\documentclass[a4 paper, 11pt]{amsart}
\usepackage{geometry}
\usepackage{float, amssymb, amsmath, amsthm, amstext, amsbsy}
\usepackage{caption}
\usepackage{subcaption}
\usepackage{epstopdf}
\usepackage{xspace}
\usepackage{graphicx, color}
\usepackage{cite}
\definecolor{darkblue}{rgb}{0.06,0.008,0.471}
\definecolor{darkgreen}{rgb}{0.008,0.871,0.067}
\usepackage[colorlinks,pdfusetitle,urlcolor=darkblue,citecolor=darkblue,linkcolor=darkblue,bookmarksnumbered,plainpages=false]{hyperref}
\usepackage[usenames,dvipsnames,table]{xcolor}
\usepackage{algorithm}
\usepackage{amsgen,amsmath, amsthm, amstext,amsbsy,tikz,amssymb,tkz-linknodes, tkz-graph, tkz-berge}
\usepackage{algpseudocode}
\renewcommand{\algorithmicrequire}{\textbf{Input:}}
\renewcommand{\algorithmicensure}{\textbf{Output:}}
\newcommand{\algorithmicbreak}{\textbf{break}}
\newcommand{\BREAK}{\State \algorithmicbreak}
\usepackage{slashbox}
\usepackage{multirow}
\usepackage{rotating, pdflscape}

\newcommand{\CNi}{\textnormal{CN}^{(i)}}
\newcommand{\CNoN}{\textnormal{CN}^{(1,2)}}
\newcommand{\Ne}{\textnormal{N}}
\newcommand{\DT}{d_{td}}
\usepackage[utf8]{inputenc}

\begin{document}

\title[Improving detection of influential nodes in complex networks]{Improving detection of influential nodes in complex networks}
\author[A. Sheikhahmadi]{Amir Sheikhahmadi\textsuperscript{\lowercase{\textsf{1,$\ast$}}}} 
\author[M. A. Nematbakhsh]{Mohammad A. Nematbakhsh\textsuperscript{\textsf{1}}} 
\thanks{\textsuperscript{1}Department of Computer Engineering, University of Isfahan, Isfahan, Iran. 
 }
\author[A. Shokrollahi]{Arman Shokrollahi\textsuperscript{\lowercase{\textsf{2,3,$\ast$}}}}
\thanks{\textsuperscript{\lowercase{\textsf{2}}}Department of Mathematics and Computer Science, West Virginia University, Morgantown, WV 26506, USA}
\thanks{\textsuperscript{\textsf{3}}Center for Artificial Intelligence and Network Science, Sanandaj, Kurdistan, Iran.}
\thanks{\textsuperscript{\lowercase{\textsf{$\ast$}}}These authors contributed equally to this work.}
\date{March 28, 2015}

\begin{abstract}
Recently an increasing amount of research is devoted to the question of how the most influential nodes (seeds) can be found effectively in a complex network. There are a number of measures proposed for this purpose, for instance, high-degree centrality measure reflects the importance of the network topology and has a reasonable runtime performance to find a set of nodes with highest degree, but they do not have a satisfactory dissemination potentiality in the network due to having many common neighbors ($\textnormal{CN}^{(1)}$) and common neighbors of neighbors ($\textnormal{CN}^{(2)}$). This flaw holds in other measures as well. In this paper, we compare high-degree centrality measure with other well-known measures using ten datasets in order to find a proportion for the common seeds in the seed sets obtained by them. We, thereof, propose an improved high-degree centrality measure (named \textit{DegreeDistance}) and improve it to enhance accuracy in two phases, FIDD and SIDD, by putting a threshold on the number of common neighbors of already-selected seed nodes and a non-seed node which is under investigation to be selected as a seed as well as considering the influence score of seed nodes directly or through their common neighbors over the non-seed node. To evaluate the accuracy and runtime performance of DegreeDistance, FIDD, and SIDD, they are applied to eight large-scale networks and it finally turns out that SIDD dramatically outperforms other well-known measures and evinces comparatively more accurate performance in identifying the most influential nodes.
\end{abstract}
\maketitle
\small{\tableofcontents}

\section{Introduction} \label{Introduction}

Identifying the most influential nodes is a pivotal challenge and is of high importance due to its efficacious applications in complex networks, such as proliferation or ceasing the influence over social and economic networks or giving publicity to a product, organization, or venture \cite{estevez2007selecting, kimura2007extracting, kimura2010extracting, zhang2010identifying}, prevention and control of infectious diseases, understanding the function of the human brain and mental disorders \cite{moussa2011changes, sporns2013structure}, ranking web pages properly in search engines results \cite{langville2011google, page1999pagerank}, further analysis of the most enriched processes in biological systems and therapeutic targets \cite{arodz2015identifying}.  Typically in social networks where the number of users is considerably increasing, one of the goals is maximizing or minimizing the spread of influence through influential nodes. The compulsive, entertaining environments of these networks and the wide diversity of services these systems provide, are making them a proper place for amusement, training, propaganda, etc \cite{heidemann2012online}. Everyday, we see a huge amount of goods and products advertisements, campaign people ads, and etc over these networks. Accepting an advertisement by a user and sharing it with friends and again friends with their friends actively publicize it and facilitates propagation \cite{dinh2012cheap, hinz2014new, iyengar2011opinion}. It basically takes advantage of users to advertise products without too much sustained efforts rather than direct interaction which is very costly. On the other hand, the result of this process may be more efficient if friends have confidence in one another \cite{cheung2009credibility, chevalier2006effect, koh2010online, park2007effect, yang2010study}. This interactive marketing technique is known as ``viral marketing" which induces social networking services and other technologies to pass along a marketing message by finding and convincing the most influential individuals \cite{dinh2012cheap, hinz2014new, iyengar2011opinion, cheung2009credibility, chevalier2006effect, koh2010online, park2007effect, richardson2002mining, probst2013will}. Shortly after, some immediate questions come up like what is the influential node? and how can they be identified?
Indeed it is not practically feasible to select all these typical nodes to start propagation due to  a shortage of funds and time-consuming, expensive process. Accordingly, the problem is to find an optimal subset of nodes within the network that are able to spread the influence and information as efficient and effective as possible. Previous literature address the maximization problem as ``maximizing the spread of influence" \cite{kempe2003maximizing, chen2009efficient}. \\

Any complex network can be modeled as a directed or undirected network (or graph) consisting of nodes (vertices) and links (edges). Due to conspicuous lack of information about nodes in some complex networks (e.g. social networks), a fairly large amount of scientific studies have considered the structural parameters \cite{chen2012identifying, gao2013modified, wei2013identifying, freeman1979centrality, yang2010study, burt2009structural, bonacich2001eigenvector}. Then, nodes have been ranked based on the topology of the network and the location of each node in the network. In these approaches, nodes have been evaluated based on measures such as high-degree (or simply degree), betweenness, closeness, etc, and those with the highest/lowest measure have been taken as influential nodes (seeds) to start any desired propagation activities over the network. In this paper, we first scrutinize these measures and figure out a rate of intersection of the seed sets obtained by these measures. Another noteworthy observation is that if seeds in these seed sets are not identical, they are very close to one another so that they are either neighbors or neighbors of neighbors of each other. So, we perceive that the neighborhood overlapping of seeds of different seed sets obtained by these measures is prominent. Hence, these seed sets influence almost the same collection of nodes in the network. Figure \ref{figure1a} displays a small network and, as we can see, nodes $v_1, v_2, v_6, v_7$ show high-degree centrality which are adjacent to each other, however by choosing $v_1$ and $v_{14}$ which are in an appropriate distance of each other, we can achieve a more effective propagation.

\begin{center}
\begin{figure}[H]
\centering
\begin{tikzpicture}[-,thick, >=stealth']
\tikzstyle{every node}=[circle, draw, fill=red!30,inner sep=0pt, minimum width=5pt]
\draw (0,0) node (15) [label=below:$v_{15}$] {};
\draw (2,0) node (17) [label=below:$v_{17}$] {};
\draw (-1,1) node (16) [label=left:$v_{16}$] {};
\draw (.5,1) node (12) [label=right:$v_{12}$] {};
\draw (0.5,2) node (14) [label=above:$v_{14}$] {};
\draw (-1,2) node (13) [label=left:$v_{13}$] {};
\draw (0,3) node (11) [label=above:$v_{11}$] {};
\draw (2,2) node (10) [label=below:$v_{10}$] {};
\draw (1,4) node (18) [label=below:$v_{18}$] {};
\draw (2,5) node (6) [label=right:$v_{6}$] {};
\draw (4,3) node (7) [label=above:$v_{7}$] {};
\draw (2,6) node (19) [label=left:$v_{19}$] {};
\draw (4,2) node (9) [label=below:$v_{9}$] {};
\draw (6,2) node (8) [label=below:$v_{8}$] {};
\draw (7,3) node (2) [label=right:$v_{2}$] {};
\draw (6,3) node (1) [label=below:$v_{1}$] {};
\draw (7,4) node (3) [label=right:$v_{3}$] {};
\draw (6,5) node (4) [label=right:$v_{4}$] {};
\draw (5,5) node (5) [label=left:$v_{5}$] {};
\tikzset{every node/.style={fill=white}}
\draw (10) --  (7) -- (6) -- (10) --  (14) -- (11);
\draw (7) -- (1) -- (2);
\draw (7) -- (8) -- (2) -- (3) -- (1) -- (6) -- (19);
\draw (18) -- (6);
\draw (14) -- (13);
\draw (14) -- (12) -- (17);
\draw (12) -- (15);
\draw (12) -- (16);
\draw (1) -- (7);
\draw (1) -- (5);
\draw (1) -- (4);
\draw (7) -- (9);
\end{tikzpicture}
\caption{A sample network which demonstrates that we get better propagation if the seed nodes ($v_1$ and $v_{14}$) are chosen in an appropriate distance of each other.} \label{figure1a}
\end{figure}
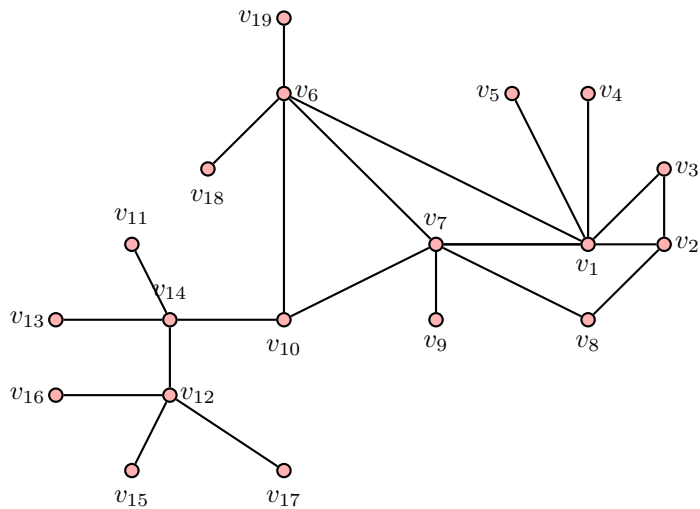
\end{center}

Hereinafter, we use the following concepts and notations throughout the paper: The \emph{distance} between two nodes $v$ and $w$, denoted by $d(v,w)$, is the length of a shortest path between them. We say that a node $w$ is an \emph{$i$-th neighbor} ($i \in \mathbb{Z}^+$) of nodes $v_1, v_2, \ldots , v_r$, $r \geq 1$, if $d(v_1, w) = d(v_2, w) = \cdots = d(v_r , w) = i$. Let $\Ne^{(i)} (v_1, v_2, \ldots, v_r)$ denote the family of all $i$-th neighbors of nodes $v_1, v_2, \ldots, v_r$, and $\Ne^{(i)}$ if nodes are not specified. If $A = \{v_1, v_2, \ldots, v_r \}$, we use the short notation $\Ne^{(i)} (A)$. In some network science and graph theory texts, $\Ne^{(1)} (v)$ and $\Ne^{(2)}(v)$ are referred to as neighbors of $v$ and neighbors of neighbors (second order contiguity) of $v$, respectively. A node $z$ is said to be an \emph{$i$-th common neighbor} of nodes $v_1, v_2, \ldots, v_r$, $r \geq 1$, if $z \in \bigcap_{h=1}^{r} \Ne^{(i)} (v_h)$. We denote the set of all $i$-th common neighbors of nodes $v_1, v_2, \ldots, v_r$ by $\CNi (v_1, v_2, \ldots, v_r)$, and $\CNi$ if $v_h$'s ($h=1,2,\ldots, r$) are not specified.  We define $\CNoN = \textnormal{CN}^{(1)} \cup \textnormal{CN}^{(2)}$. A node $w$ is said to be in \emph{distance threshold}, $\DT$, from $v$ if $w \in \Ne^{(r)} (v)$ for some $r \geq \DT$. 
A node $w$ is said to be \emph{unique} in sets $X_1, X_2, \ldots, X_r$, $r \geq 1$, if there exists one and only one $h \in \{ 1,2,\ldots, r \}$ such that $w \in X_h$. Lastly, let $k$ be the seed set size. \\
For example, in Figure \ref{figure1a}, $\Ne^{(2)} (v_7) = \{ v_2, v_3, v_4, v_5, v_{14}, v_{18}, v_{19} \}$; $v_2 \in \textnormal{CN}^{(3)} (v_{10}, v_{18}, v_{19})$; $v_1$ is a unique node in $\Ne^{(1)} (v_6)$, $\Ne^{(2)} (v_7)$, and $\textnormal{CN}^{(3)} (v_{10}, v_{19})$ because $v_1 \in \Ne^{(1)} (v_6)$ only; if we want to take a node in distance threshold $\DT = 2$ from $v_{13}$, we can choose any node in the network but $v_{14}$, similarly there is no node in distance threshold $\DT = 4$ of $v_{10}$. \\

In this study, we first investigate structural measures including high-degree, betweenness, closeness, eigenvector, PageRank, LeaderRank, and $k$-shell to show that regardless of the type of the measure and performance variety, the seed sets they produce have many seeds in common. We then verify that these structural measures usually search and select the nodes in the least distance within the network. Finally, we propose a method (named \emph{DegreeDistance}) to find the most influential nodes by reforming high-degree centrality measure. Roughly speaking, we discuss and present: (1) \textit{DegreeDistance}: an improved high-degree centrality measure in order to select the seed set, (2) \textit{FIDD (First Improvement of DegreeDistance)}:  an improvement of DegreeDistance by analyzing the number of common neighbors of seeds up to a distance threshold $\DT \in \{2,3\}$, (3) \textit{SIDD (Second Improvement of DegreeDistance)}:  an improvement of FIDD by applying the influence score of the already-selected nodes in the seed set and their neighbors over a new potential node which is under investigation to be selected as a seed.

The main advantage of our proposed methods is greater performance in maximizing influence propagation with reasonable running time. \\

 The rest of this paper is organized as follows: Section \ref{structural_approaches} briefly overviews well-known structural measures which build the basis of our discussions. In Section \ref{main_results}, we present the steps of \textit{DegreeDistance} which is similar to high-degree centrality in spirit, and its improvements, FIDD and SIDD, to effectively and efficiently select the most influential nodes. In Section \ref{evaluation}, we compare our methods with other measures, and in the last section, we summarize the main conclusions and suggest possible future directions.

\section{Structural measures}  \label{structural_approaches}

The problem of identifying the most influential nodes in order to spread information over complex networks has been already studied in \cite{yang2010study, richardson2002mining, probst2013will, kempe2003maximizing, chen2009efficient, chen2012identifying, gao2013modified, wei2013identifying, freeman1979centrality, burt2009structural,  bonacich2001eigenvector, katsaros2015detecting, liu2015new}. There are well-known measures that mostly deal with the location of nodes in the network. We use some of them to show that their seed sets contain partially the same seeds, and the seeds in a seed set have a significantly large amount of $\CNoN$. We also utilize the best measures among them to test the performance of our proposed methods. In the following, we briefly sketch them.

\subsection{High-degree centrality} \label{high-degree_method}
In this method, simply the nodes with the highest degree in the network should be marked as seeds. The reason behind this strategy is that these nodes can influence more nodes effectively due to having the greatest number of neighbors \cite[Ch. 3]{golbeck2013analyzing}. High-degree centrality has been considered as a measure to study complex networks and the importance of nodes in (un)weighted networks \cite{chen2012identifying, gao2013modified, wei2013identifying}. \\

 L. Katz \cite{katz1953new} developed this concept and introduced \textit{Katz centrality} to measure the degree of influence of a node which takes into account the total number of walks. Each connection with distance $j$ will be penalized by $\beta^j$ where $0 \leq \beta \leq 1$. The formula to compute this measure is as follows,
\begin{equation} \label{eq1}
C_i^{\textnormal{Katz}} = e_i^T \left(\sum_{j=1}^{\infty} (\beta \mathbf{A})^j \right) \mathbf{I},
\end{equation}
where $e_i$ is a column vector whose entries are all zero except the $i$-th entry which is 1, and $\mathbf{I}$ is the identity matrix. The disadvantage of using high-degree centrality measure is that it considers a node locally, i.e. based on its location, and in graphs with multiple components, the seeds are likely to be selected only from a big component.

\subsection{Closeness centrality} \label{closeness_method}
The \textit{farness} of a node $u$ is the sum of the distances of $u$ to all other nodes, and its closeness is the reciprocal of the farness. Hence, the closeness can be interpreted as a measure indicating how long it will take to spread information from a node $u$ to all other nodes sequentially, another words, $u$ is taken as an influential (central) node by the closeness strategy if its total distance to all other nodes is lowest. These nodes have greater influence due to the least number of intermediaries. This centrality measure can be computed by counting the shortest paths, and the following is one of the well-known expressions that is attributed to sociologist L. Freeman \cite{freeman1979centrality},
\begin{equation}
\label{eq2}
C_i^{\textnormal{CLO}} = e_i^T \mathbf{S} \mathbf{I},
\end{equation}
where $\mathbf{S}$ is the matrix whose $(i,j)$-th entry represents the length of a shortest path from node $i$ to node $j$. The closeness measure needs to travel over the whole network, and clearly it is time-consuming and inappropriate for large-scale networks.

\subsection{Betweenness centrality} \label{betweenness_method}
By this indicator, influential nodes are those that are visited by the largest number of shortest paths from all nodes to all others within the network. L. Freeman \cite{freeman1979centrality} has introduced the expression below to compute this centrality,
\begin{equation}
\label{eq3}
C_i^{\textnormal{BET}} = \sum_{j \neq r \neq i}^{} \frac{g_{jr}(i)}{g_{jr}},
\end{equation}
where $g_{jr}$ is the number of shortest paths between nodes $j$ and $r$, and $g_{jr} (i)$ is the number of shortest paths between $j$ and $r$ passing through the node $i$.

The nodes with the highest betweenness are sometimes called \textit{bottlenecks} \cite{yu2007importance}, or \textit{intermediaries} \cite{scott2011sage}, or  \textit{structural holes} \cite{burt2009structural}.

\subsection{Eigenvector centrality} \label{eigenvector_method}
This measure is closely related to Katz centrality and was introduced first by P. Bonacich \cite{bonacich2001eigenvector}. It tries to find the influence of a node by assigning a score to every node based on the adjacency of that node to high-scoring nodes.

\subsection{PageRank} \label{PageRank_method}
PageRank is an algorithm which is used in Google search engine to rank web pages \cite{brin1998anatomy}. A web page linking to more important web pages has higher rank. Thus, a page with fewer neighbors might have a higher PageRank than another page with more neighbors. S. Fortunato et al. \cite{fortunato2005make} and J. Heidemann et al. \cite{heidemann2010identifying} separately used this centrality measure to rank nodes in social networks.

\subsection{DegreeDiscount centrality} \label{Degree_discount_method}
In 2009, W. Chen et al. \cite{chen2009efficient} proposed the DegreeDiscount heuristic algorithm. When a node is selected as a seed, another node with highest degree can be potentially selected as a new seed, but the edge between these two should not be counted towards its degree \cite[Ch. 4]{aggarwal2011introduction}. Another words, if a node $u$ has degree $d_u$, and $d^\prime_u$ of them are already selected as seeds, we need to discount $d(u)$ by $2d^\prime_u + (d_u - d^\prime_u)d^\prime_u p$, where $p$ is a small propagation probability. This model does not maximize the total information flow in the network.

\subsection{LeaderRank}
In 2011, L. L\"u et al. \cite{lu2011leaders} proposed a variant of PageRank known as LeaderRank.  Weighted LeaderRank is a slightly improved version of LeaderRank \cite{li2014identifying}.

\subsection{$k$-shell decomposition}
M. Kitsak  et al. \cite{kitsak2010identification} presented this measure which basically deals with the location of nodes in the network and assigns a $k_S$ index to each node. Nodes with high index are located in the innermost network core and those with low index are at the periphery of the network.

\subsection{Greedy algorithm}
This algorithm introduced by D. Kempe et al. \cite{kempe2003maximizing}.
An initial seed set, $S$, is considered and in each step of the algorithm a single node, $v$, is being added to $S$ so that $S \cup \{ v \}$  maximizes the spread of influence and activates a larger number of nodes in the network. This process iteratively continues until the top $k$ nodes are chosen, i.e. $|S| = k$.

\section{Our centrality measure, \textit{DegreeDistance}, and its improvements} \label{main_results}

In this section, we first discuss this matter that the well-known measures, mentioned in the preceding section, select almost the same seed set, and then find the rate of similarity between neighbors and neighbors of neighbors of seeds of the seed set obtained by any of the measures (i.e. $\textnormal{CN}^{(1)}$ and $\textnormal{CN}^{(2)}$ of seed nodes obtained by a particular measure). Based on this argument, we build a new seed set by exclusion of neighbors of seed nodes up to a specific distance, so seeds will be in distance threshold, $\DT$, from each other, and we propose a technique to improve identifying the most influential nodes.

\subsection{Common seeds of different seed sets}

The main question here is how many seeds do the seed sets obtained by the mentioned measures have in common? To be more clear, we can find the number of common seeds obtained by, for example, high-degree and closeness, or closeness and PageRank, etc. We also address the total cardinality of $\CNoN$ of seeds in a seed set.

\begin{center}
\begin{figure}
\centering
\includegraphics[width = 4.5in]{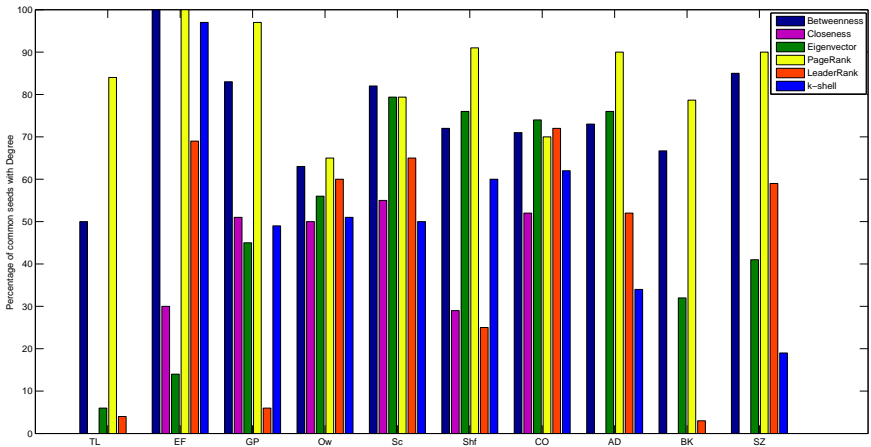}
\caption{The percentage of common seeds between high-degree seed set and betweenness, closeness, eigenvector, PageRank, LeaderRank, and $k$-shell seed sets are shown in dark blue, magenta, green, yellow, red,  and light blue, respectively. Here $k=100$.}
\label{fig23}
\end{figure}
\end{center}

To find out the number of common seeds, we take out the first $k$ seeds using each measure, where $k \in \{25, 50, 75, 100 \}$ in our investigation, and apply the following formula,
\begin{equation}
\label{eq4}
COM(S_1, S_2) = \frac{|S_1 \cap S_2|}{k} \cdot 100,
\end{equation}
where $S_1$ and $S_2$ are two seed sets obtained from two arbitrary centrality measures.  To investigate this type of overlapping, we use the first ten datasets described in Table \ref{table1}. All the datasets are taken from KONECT except NetHEPT which is a scientific collaboration network taken from the High Energy Physics - Theory citations from arXiv. Since we are particularly interested in high-degree centrality measure, we have examined the number of its common seeds with other measures' seed sets in Figure \ref{fig23}.

\begin{table}
\caption{The list of the real-world datasets used in this paper. Order and size are the number of nodes and edges, resp.} \label{table1}
\begin{center}
\rowcolors{1}{white}{pink!30}
\begin{tabular}{|llcccc|}
\hline
\textsc{Dataset}     & \textsc{Type}   & \textsc{Order} & \textsc{Size} & \textsc{Avg Degree} & \textsc{Max Degree}  \\ \hline\hline
Twitter lists (TL)  & directed & 23370 & 33101 & 2.8328 & 239   \\
Facebook-NIPS (EF)  & directed & 2888 & 2981 & 2.0644 & 769  \\
Google+ (GP) & undirected & 23628 & 39242 & 3.3217 & 2771 \\
Facebook wall posts (Ow) & directed & 46952 & 876993 & 37.357 & 2696  \\
Catster (Sc) & undirected & 149700 & 5449275 & 72.803 & 80635 \\
Hamsterster friendships (Shf) & undirected & 1858 & 12534 & 3.492 & 272 \\
Wikipedia conflict (CO) & undirected & 118100 & 2917785 & 49.412& 136192 \\
Advogato (AD) & directed & 6541 & 51127 & 15.633 &	943 \\
Brightkite (BK)  & undirected & 58228 & 214078 & 7.353 & 1134 \\
Slashdot Zoo (SZ) & directed & 79120 & 515397 & 13.49028 & 2543  \\
Epinions (ES) & directed & 75879 & 508837 & 13.412 & 3079 \\
Flickr (Fl) & undirected & 105938 & 2316948 & 43.742 & 5425\\
Gowalla (GW)  & undirected & 196591 & 950327 & 9.6681 & 14730 \\
Youtube friendship (CY) & undirected & 1134890 & 2987624 & 5.2650 & 28754 \\
NetHEPT  & undirected & 15233 & 31399  & 4.12 & 64 \\
\hline
\end{tabular}
\end{center}
\end{table}

\subsection{$\CNoN$ of seeds in a seed set} \label{commonSeeds}

By computing $\CNoN$ of seeds inside a seed set, we can easily find out how topologically close they are to each other. We want to show that the seeds selected each of the mentioned measures mostly belong to $\Ne^{(i)}$, $i \leq 2$, of each other. This fact leads to wasting time and energy as well as ill-suited dissemination in complex networks. For instance, looking from the perspective of social networks, selecting seeds close to each other results in increasing persistence and intensity of a specific people in the network, based on the law of diminishing returns \cite[Ch. 7]{hirschey2008managerial}. Accordingly, we first find the rate of $\textnormal{CN}^{(1)}$ and $\textnormal{CN}^{(2)}$ (i.e. $\CNoN$) of seeds obtained by each measure. For, we first select top $k$ seeds ($k \in \{ 25, 50, 75, 100 \}$) by one of the measures, and then find the $\CNoN$ of them, put them all in $F$. We then compute the number of unique nodes, and find the rate by  $COV = 100 - \left[(unique/total)\cdot 100 \right]$. Algorithm \ref{alg:1} illustrates this procedure, and based on it, the rate of having $\CNoN$ for the seeds in different seed sets is displayed in Table \ref{table2} after we introduce DegreeDistance in Algorithm \ref{alg:2}. From the table, we can see that the DegreeDistance seeds with $\DT = 3$ have the least value of $\CNoN$, which means the seeds are in an appropriate distance of each other, and hence, they influence a larger number of unique nodes within the network, as depicted in Figure \ref{fig:comp}. To be more clear, seeds not too close to each other can influence other nodes in the network rather than influencing a specific set of nodes repeatedly, though in the continuation of the paper, we show that the value of $COV$ for seeds is not the only factor which matters and this brings some improvements into conversation.

\begin{algorithm}
\caption{Computing the rate of $\CNoN$ of $k$ seeds} \label{alg:1}
\begin{algorithmic}[1]
\Require $S_1$, $k$ \Comment{$S_1$ is the seed set}
\Ensure The percentage of $\CNoN$ of the seeds in $S_1$
\State $i \gets 0$  \Comment{$i$ is the number of selected seeds}
\State $total \gets 0$
\State $F \gets \emptyset$
\While {$i<k$}
\State $s^\prime \gets \textnormal{Top}(S_1)$
\State $S_1 \gets S_1 \setminus \{ s^\prime \}$
\State $F_i \gets \Ne^{(1)}(s^\prime) \cup \Ne^{(2)}(s^\prime) $
\State $total = total + |F_i|$
\State $i \gets i+1$
\EndWhile
\State $unique \gets |F|$
\State $COV \gets 100 - \left[(unique/total)*100 \right]$
\end{algorithmic}
\end{algorithm}

\subsection{DegreeDistance: Improved high-degree centrality measure}

As we discussed, one of the main issues with most of the widely-used measures such as high-degree, betweenness, closeness, eigenvector, PageRank, LeaderRank, and $k$-shell to select an appropriate seed set is that the seed nodes have a remarkable amount
of $\CNoN$ with one another. Therefore, due to this fact and high speed selection of seeds by high-degree centrality, the next logical step is to improve this measure in order to end up with a more effective seed set whose elements have the least value of
$\CNoN$. In our proposed method which is described in Algorithm \ref{alg:2}, we first compute the degree of each node in the network and select a node with the highest degree and add it to a predefined selection set ($Sel$). To reduce the number of elements of $\CNoN$ of the selected nodes in $Sel$, once we add a node to $Sel$,
we take a distance threshold, $\DT$, to select the next seed, namely we remove the candidacy of the neighbors of the node in distance up to $\DT$; for instance, if a node $v$ is already selected as a seed and $\DT = 3$, the nodes in $\Ne^{(1)}(v)$ and $\Ne^{(2)}(v)$ will not be checked for selecting more seeds. As a matter of fact, in social networks, we nominate a person for being a seed if its $i$-th neighbors ($i = 1,2$), who have the highest confidence in them, have the least overlapping with the $i$-th neighbors of already-selected people.

\begin{algorithm}
\caption{DegreeDistance centrality measure} \label{alg:2}
\begin{algorithmic}[1]
\Require $G$, $k$, $\DT$ \Comment{$G$ is the given network, and $\DT$ is the distance threshold}
\Ensure $S$ \Comment{seed set}
\State $S \gets \emptyset$
\State Compute degree of all nodes in $G$
\State $L \gets \textnormal{Descending list of nodes based on their degree}$
\While{$|S|<k$}
\State $s^\prime \gets \max(L)$
\State $Sel \gets \textnormal{True}$
\ForAll{$v \in S$}
\If{$d(s^\prime , v) < \DT$}
\State $Sel \gets \textnormal{False}$
\BREAK
\EndIf
\EndFor
\If{$Sel$}
\State $S \gets S \cup \{ s^\prime \}$
\EndIf
\State $L \gets L \setminus \{ s^\prime \}$
\EndWhile
\end{algorithmic}
\end{algorithm}

If $\DT = 2$, one can replace Algorithm \ref{alg:2} with Algorithm \ref{alg:3}.

\begin{algorithm}
\caption{DegreeDistance with threshold 2} \label{alg:3}
\begin{algorithmic}[1]
\Require $G$, $k$
\Ensure $S$ \Comment{seed set}
\State $S \gets \emptyset$
\State Compute degree of all nodes in $G$
\State $L \gets \textnormal{Descending list of nodes based on their degree}$
\While{$|S|<k$}
\State $s^\prime \gets \max(L)$
\State $S \gets S \cup \{ s^\prime \}$
\State $L \gets L \setminus \left\{ s^\prime \cup \Ne^{(1)}(s^\prime) \right\}$
\EndWhile
\end{algorithmic}
\end{algorithm}

In the last algorithm above, once we select a node, its neighbors will be removed from $L$, and so there exists either no path or a path of length $\geq 2$ between any two seeds. Now, it is time to show the results from Subsection \ref{commonSeeds}.

\begin{landscape}
\begin{table}
\caption{The rate of $\CNoN$ of seeds of different seed sets obtained by various measures on seven datasets. The last two columns belong to DegreeDistance (DD) with different distance threshold, $\DT = 2,3$, between seeds.} \label{table2}
\begin{center}
\small
\begin{tabular}{|lcccccccc>{\columncolor{pink!30}}c>{\columncolor{pink!30}}c|}
\hline
\textsc{Dataset}    & \textsc{Top $k$}   & \textsc{Degree} & \textsc{Betweenness} & \textsc{Closeness} & \textsc{Eigenvector} & \textsc{PageRank} & \textsc{LeaderRank} & \textsc{$k$-shell} & \textsc{DD, $\DT=2$} & \textsc{DD, $\DT=3$} \\ \hline \hline
\multirow{4}{*}{AD} & 25 & 93.44 & 93.28 & 28.00 & 93.65 & 93.35 & 93.22 & 92.11 & 81.46 & 71.31 \\
 & 50 & 96.52 & 96.42 & 14.00 & 96.59 & 96.48 & 96.17 & 95.74 & 82.24 & 72.55 \\
 & 75 & 97.57 & 97.51 & 25.68 & 97.61 & 97.56 & 97.24 & 96.97 & 84.17 & 73.51 \\
 & 100 & 98.08 & 98.05& 44.49 &	98.13 &	98.08 & 97.81 & 97.66 & 85.7 & 74.45 \\ \hline
\multirow{4}{*}{Ow} & 25 & 88.04&	86.39&	90.53&	90.35&	88.04 & 69.6 & 89.16 & 58.05 & 48.33 \\
 & 50 &  92.16& 92.16&	93.43&	94.38&	92.16 & 74.62 & 92.86 & 59.5 & 51.35 \\
 & 75 &  93.36&	93.36&	94.93&	96.04&	93.36 & 82.42 & 94.05 & 61.34 & 53.25 \\
 & 100 & 94.33&	94.33&	95.87&	96.94&	94.33 & 89.44 & 94.38 & 62.12 & 55.39 \\ \hline
\multirow{4}{*}{GP} & 25 & 82.67 & 79.12 & 74.4 & 86.49 & 78.56  & 59.27 & 87.46 & 52.88 & 41.13  \\
 & 50 & 88.79 & 87.87 & 90.06 & 93.03 & 87.76 & 86.53 & 93.61 & 54.39 & 43.19  \\
 & 75 & 91.02 & 91.6  & 93.22 & 95.36 & 90.95 & 91.09 & 95.53 & 56.98 & 45.25 \\
 & 100 & 92.94 & 93.67 & 95.38 & 96.46 & 92.77 & 92.78 & 96.48 & 58.22 & 47.89 \\ \hline
\multirow{4}{*}{TL} & 25 & 35.58 & 46.92 & 30.51 & 91.13 & 27.23 & 54.45 & 14.78 & 26.78 & 19.62 \\
 & 50 & 40.82 & 62.29 & 47.5 & 95.02 & 39.03 & 63.6 & 26.2 & 27.11 & 21.18 \\
 & 75 & 53.05 & 65.37 & 52.61 & 96.3 & 44.3 & 66.77 & 41.65 & 29.01 & 22.89 \\
 & 100 & 56.18 & 70.23 & 59.03 & 96.95 & 50.48 & 72.53 & 47.58 & 31.25 & 24.11 \\ \hline
\multirow{4}{*}{BK} & 25 & 88.26 & 88.50 & 0.00& 89.42 & 87.31 & 50.9 & 64.76 & 58.2 & 50.25 \\
 & 50 &  92.94& 93.04 & 0.00 & 92.68 & 93.10 & 60.21 & 75.7 & 59.24 & 54.18 \\
 & 75 & 94.71 & 95.00 & 0.00 & 94.32 & 94.97 & 66.07 & 80.41 & 61.33 & 57.31   \\
 & 100 & 95.20 & 95.29 & 50.00 & 94.20 & 95.27 & 70.22 & 84.02 & 62.14 & 59.55 \\ \hline
\multirow{4}{*}{Sc} & 25 &  92.77&	92.54 &	41.51&	93.26&	92.6  & 91.71 & 92.14 & 71.11 & 60.21 \\
 & 50 & 96.078&	95.81 &	52.7 &	96.3 &	95.93  & 94.29 & 95.68 & 73.25 & 63.42   \\
 & 75 & 97.3  & 96.87  & 74.03 & 97.39 & 97.16 & 95.27 &  97.06 & 76.41 & 65.25  \\
 & 100 &  97.87 & 97.53& 93.92 & 97.93 & 97.79  & 96.3 & 97.81 & 78.52 & 68.19  \\ \hline
\multirow{4}{*}{SZ} & 25 & 81.55 & 81.07 &	0.00 & 80.99 & 80.95  & 87.59 & 87.33 & 68.45 & 59.94 \\
 & 50 & 88.46 &	88.09 &	0.00 & 87.84 & 88.20 & 92. 49 & 92.61 & 71.45 & 63.66 \\
 & 75 & 91.21 & 90.75 &	0.00 & 90.50 &	90.89  & 94.36 & 94.64 & 73.67 & 65.15 \\
 & 100 & 96.17& 96.07 & 50.00 & 96.01 & 96.01  & 95.52 & 95.84 & 76.18 & 68.38  \\ \hline
\end{tabular}
\end{center}
\end{table}

\begin{table}
\caption{A sample of information about the cardinality of $\CNoN$ for different values of $k$ on the AD dataset.} \label{table3}
\begin{center}
\small
\begin{tabular}{|cclccccccc>{\columncolor{pink!30}}c>{\columncolor{pink!30}}c|}
\hline
\textsc{Dataset}     & \textsc{Top $k$}   & \textsc{} & \textsc{Degree} & \textsc{Betweenness} & \textsc{Closeness} & \textsc{Eigenvector} & \textsc{PageRank} & \textsc{LeaderRank} & \textsc{$k$-shell} & \textsc{DD, $\DT=2$} & \textsc{DD, $\DT=3$} \\ \hline\hline
\multirow{8}{*}{AD} & \multirow{2}{*}{25}  &  \textit{total:} & 74487 & 73031 & 50 & 76421 & 73497 & 57040 & 68429 & 27650 & 18545  \\ 
   &   &   \textit{unique:} & 4887 & 4905 & 36 & 4853 & 4890  & 4501 & 4641 & 5125 & 5320  \\ \cline{2-11}
 & \multirow{2}{*}{50}  &\textit{total:} & 141671 & 138247 & 100 & 143725 & 140506  & 109983  & 124113 & 29780 & 20200  \\ 
    &   &   \textit{unique:} & 4933 & 4944 & 86 & 4904 & 4939 & 4679 &  4749 & 5289 & 5545  \\ \cline{2-11}
 & \multirow{2}{*}{75}  &  \textit{total:} & 204171 & 200079 & 148 & 207220 & 204059  & 158952 &  175365 & 33750 & 22350 \\ 
    &   &   \textit{unique:} & 4966 & 4973 & 110 & 4946 & 4969  & 4808 & 4842 & 5340 & 5920 \\
    \cline{2-11}
     & \multirow{2}{*}{100}  &  \textit{total:} & 260044 & 256520 & 236 & 266417 & 260010  & 221245 &  209872 & 37890 & 24321 \\ 
    &   &   \textit{unique:} & 4990 & 5010 & 131 & 4982 & 4992  & 4845 & 4911 & 5421 & 6213  \\ \hline
\end{tabular}
\end{center}
\end{table}
\end{landscape}

To clarify how to get the values of $COV$ in Table \ref{table2}, as a sample case, the detailed information about the cardinality of $\CNoN$ of top $k$ seeds of the AD dataset is displayed in Table \ref{table3}. Closeness seeds apparently have the least values, this is because there are heterogeneous components in the network and the tendency of this measure to small components.

\subsection{FIDD using $\textnormal{CN}^{(1)}$}

In our proposed centrality measure (i.e. DegreeDistance), if a high-degree node is selected as a seed, we then avoid selecting its neighbors up to the $\DT$ which yields an increase of spreading.
In this way, despite the location diversity of the selected nodes,
we may practically remove nodes that have a highly influential neighbor in the seed set, though their connection might be weak. For example, in Figure \ref{exampleFIDD}, the node $w_1$ with highest degree is chosen as a seed, and if the distance threshold is $\DT = 2$, the nodes in $\Ne^{(1)} (w_1)$ are practically put aside and the next seed will be $w_9$. Therefore, we see that the high degree node $w_2$ is removed and since there is only one path between $w_1$ and $w_2$, the subsequent nodes of $w_2$ will never get the chance of being influenced.

\begin{center}
\begin{figure}[H]
\centering
\begin{tikzpicture}[-,thick, >=stealth']
\tikzstyle{every node}=[circle, draw, fill=red!30,inner sep=0pt, minimum width=5pt]
\draw (0,1) node (1) [label=above:$w_1$] {};
\draw (1,1) node (2) [label=above:$w_2$] {};
\draw (-1,0) node (3) [label=left:$w_3$] {};
\draw (-1,1) node (4) [label=left:$w_4$] {};
\draw (-1,2) node (5) [label=below:$w_5$] {};
\draw (2,2) node (6) [label=right:$w_6$] {};
\draw (2,1) node (7) [label=right:$w_7$] {};
\draw (2,0) node (8) [label=right:$w_8$] {};
\draw (-2,2) node (9) [label=below:$w_9$] {};
\draw (-3,2) node (10) [label=left:$w_{10}$] {};
\tikzset{every node/.style={fill=white}}
\draw (1) --  (2) -- (6);
\draw (2) -- (7);
\draw (2) -- (8);
\draw (1) -- (3) -- (4) -- (1);
\draw (1) -- (5) -- (9) -- (10);
\end{tikzpicture}
\caption{DegreeDistance may remove neighbors of a seed which exert a powerful influence. By choosing $w_1$ and $\DT = 2$, the node $w_2$ will be removed. We present FIDD to overcome this drawback.} \label{exampleFIDD}
\end{figure}
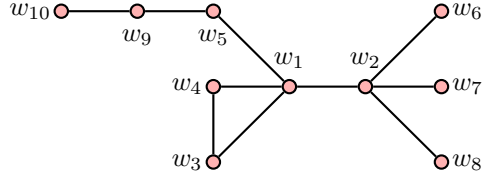
\end{center}

Based on the argument above, to nominate a new node (with highest degree among non-seed nodes) to be a seed, we need to evaluate $|\textnormal{CN}^{(1)}|$ of seed nodes and the node which is in question. If it falls below a threshold $\theta$, the node can be chosen as a seed, otherwise the influence is more likely to be easily propagated through these common neighbors, and therefore we do not select the node. This improvement is presented in Algorithm \ref{alg:4}.

\begin{algorithm}
\caption{FIDD} \label{alg:4}
\begin{algorithmic}[1]
\Require $G$, $k$, $\DT$, $\theta$ \Comment{$\DT \in \{ 2,3 \}$, and $\theta$ is the threshold for $| \textnormal{CN}^{(1)}|$}
\Ensure $S$
\State $S \gets \emptyset$
\State $L \gets \textnormal{Descending list of nodes based on their degree}$
\While{$|S|<k$}
\State $s^\prime \gets \max(L)$
\State $L \gets L \setminus \{ s^\prime \}$
\State $Sel \gets \textnormal{True}$
\ForAll{$v \in S$}
\If{$d(s^\prime , v) < \DT$}
\State $No \gets | \textnormal{CN}^{(1)} (s^\prime , v)|$
\If{$No \geq \theta$}
\State $Sel \gets \textnormal{False}$
\BREAK
\EndIf
\EndIf
\EndFor
\If{$Sel$}
\State $S \gets S \cup \{ s^\prime \}$
\EndIf
\EndWhile
\end{algorithmic}
\end{algorithm}

\subsection{SIDD using $\textnormal{CN}^{(1)}$ and the influence of seeds and their neighbors}

The point missing in the last algorithm above is that how much may a non-seed node be influenced by seed nodes and their neighbors? In this regard, we present Algorithm \ref{alg:5}.

\begin{algorithm}
\caption{SIDD} \label{alg:5}
\begin{algorithmic}[1]
\Require $D$, $k$, $\DT$, $\theta$ \Comment{$\DT \in \{ 2,3 \}$, and $\theta$ is the threshold for $| \textnormal{CN}^{(1)}|$}
\Ensure $S$
\State $S \gets \emptyset$
\State $L \gets \textnormal{Descending list of nodes based on their degree}$
\While{$|S|<k$}
\State $s^\prime \gets \max(L)$
\State $L \gets L \setminus \{ s^\prime \}$
\State $Sel \gets \textnormal{True}$
\State $\textnormal{inf} \gets 0$
\ForAll{$v \in S$}
\If{$d(s^\prime , v) < \DT $}
\State $No \gets | \textnormal{CN}^{(1)} (s^\prime , v)|$
\State $\textnormal{inf} \gets \mathbb{P}(v,s^\prime) + \sum_{w\in \textnormal{CN}^{(1)}(s^\prime, v)} \left( \mathbb{P}(v,w) * \mathbb{P}(w,s^\prime) \right)$
\If{$ No \geq \theta \And \textnormal{inf} \geq \beta$}
\State $Sel \gets \textnormal{False}$
\BREAK
\EndIf
\EndIf
\EndFor
\If{$Sel$}
\State $S \gets S \cup \{ s^\prime \}$
\EndIf
\EndWhile
\end{algorithmic}
\end{algorithm}

In SIDD measure, to determine whether ir not a new node,  $s^\prime$,  with highest degree should be selected as a seed, we add one more condition to FIDD which is the influence score and can be computed via the following expression,
\begin{equation}\label{eq5}
 \textnormal{inf} = \mathbb{P}(v,s^\prime) + \sum_{w \in \textnormal{CN}^{(1)}(s^\prime, v) } \Big( \mathbb{P}(v,w) \cdot \mathbb{P}(w,s^\prime) \Big).
 \end{equation}
Applying this expression, the activation probability of the in-question node, $s^\prime$, by a seed node $v$ such that $d(s^\prime , v) < \DT$  through nodes $w \in \textnormal{CN}^{(1)} (s^\prime , v)$, can be determined. If this score is large enough, we can remove $s^\prime$ and give the chance of being a seed to another node which has little possibility to be influenced by seed nodes directly or through their neighbors.

\section{Evaluation and experimental results}
\label{evaluation}

In this section,  we assess the rate of having $\CNoN$ of DegreeDistance seeds and the rate of the number of seeds that DegreeDistance seed set and other measures' have in common. We also the runtime performance and spread ability of influence by DegreeDistance, FIDD, and SIDD seeds,
then compare them with some other well-known measures. The proposed measures in this paper can be applied to any complex networks, albeit here we have mostly conducted the experiments on social networks and networks of this sort.

\begin{center}
  \begin{figure}[H]
  \begin{center}
  \begin{tabular}{lcc}
      \begin{subfigure}[h]{2in}
      \centering
      \includegraphics[width=2in]{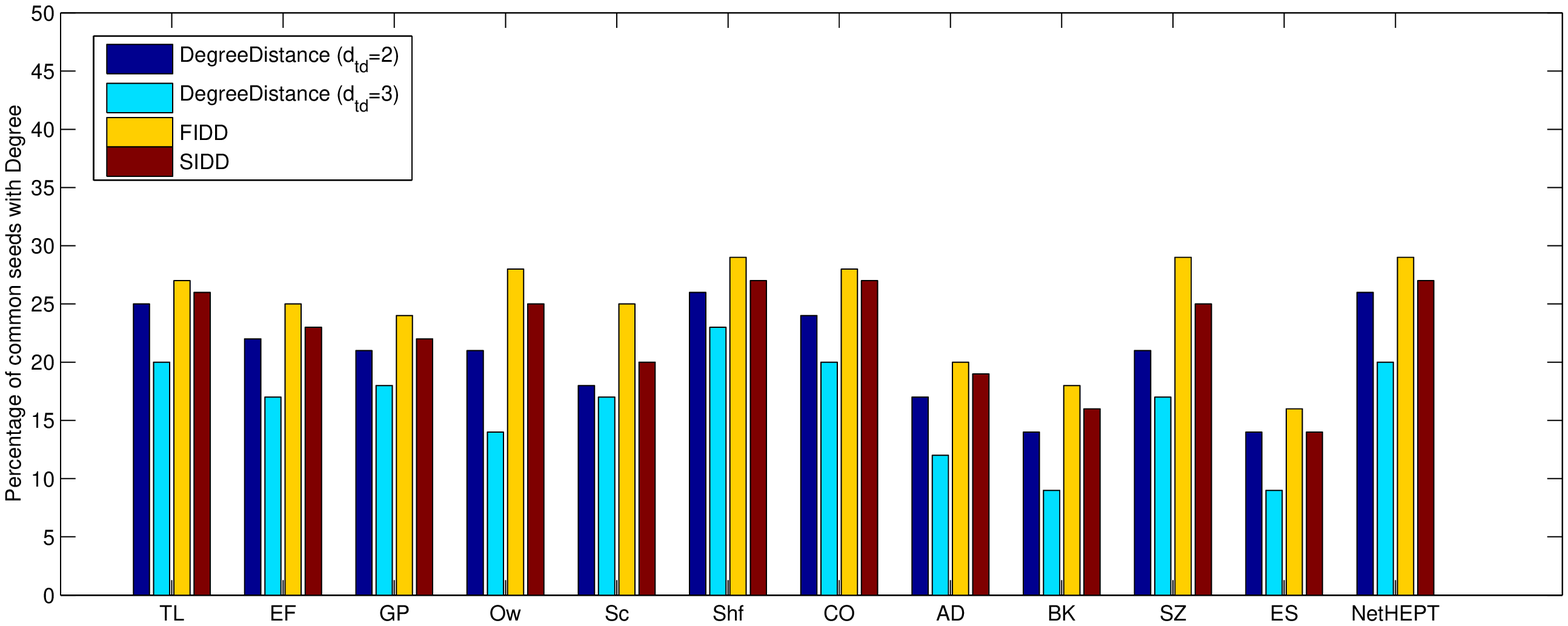}
      \caption{}
      \end{subfigure}
      &
      \begin{subfigure}[h]{2in}
      \centering
      \includegraphics[width=2in]{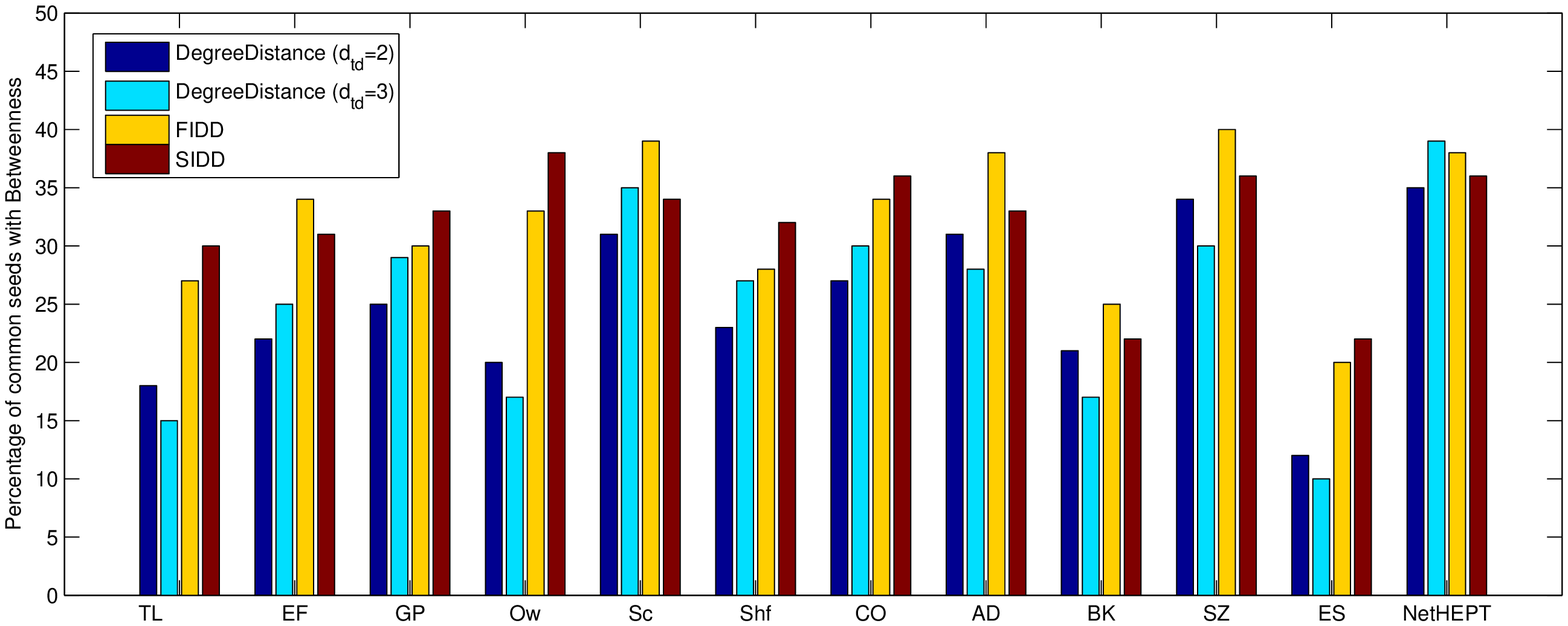}
      \caption{}
      \end{subfigure}   
      &
      \begin{subfigure}[h]{2.1in}
          \centering
          \includegraphics[width=2in, height=2.2cm]{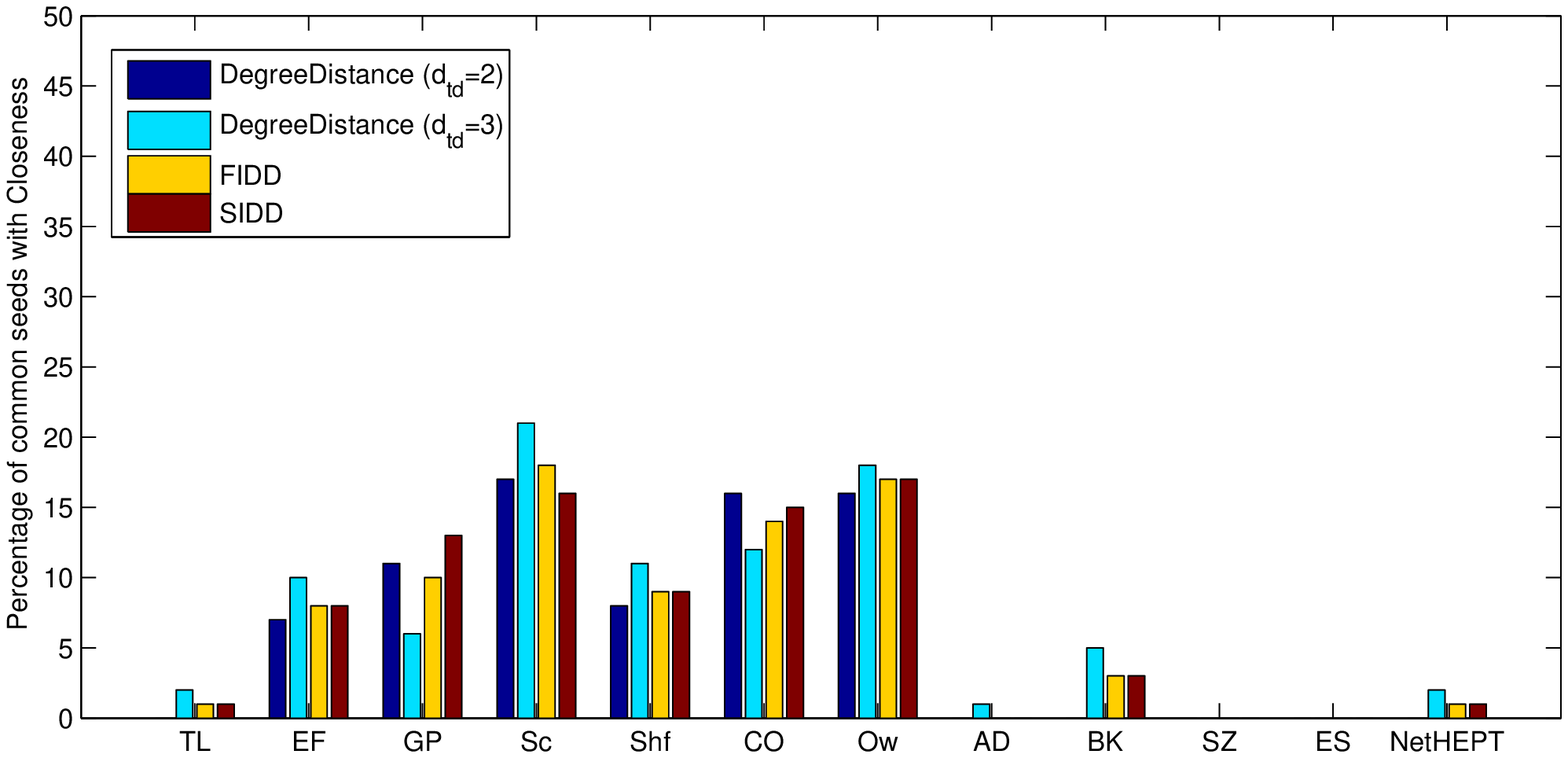}
          \caption{}
       \end{subfigure}  \\      
      \begin{subfigure}[h]{2in}
          \centering
      \includegraphics[width=2in, height=2.4cm]{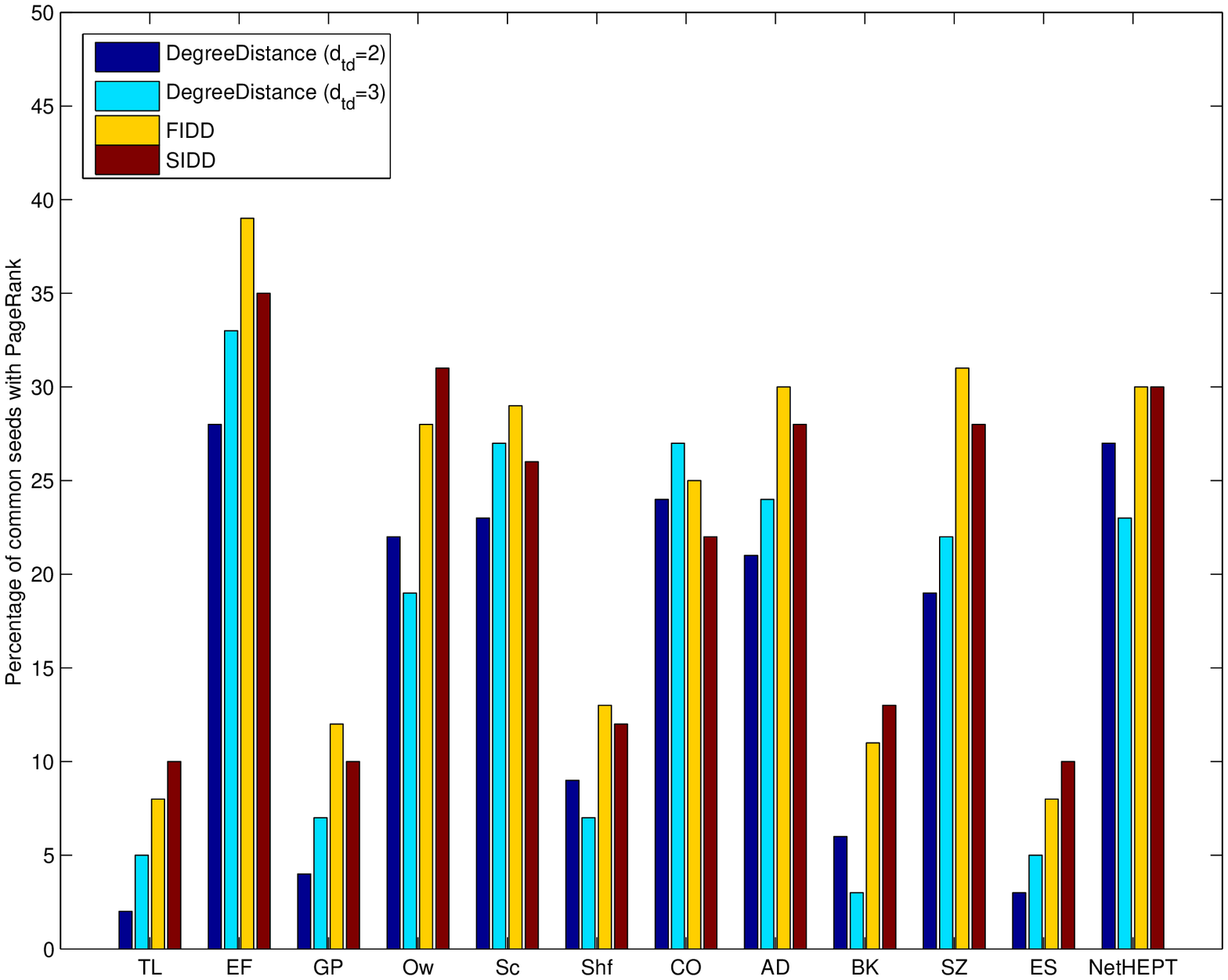}
      \caption{}
      \end{subfigure} 
   &
       \begin{subfigure}[h]{2in}
              \centering
              \includegraphics[width=2in]{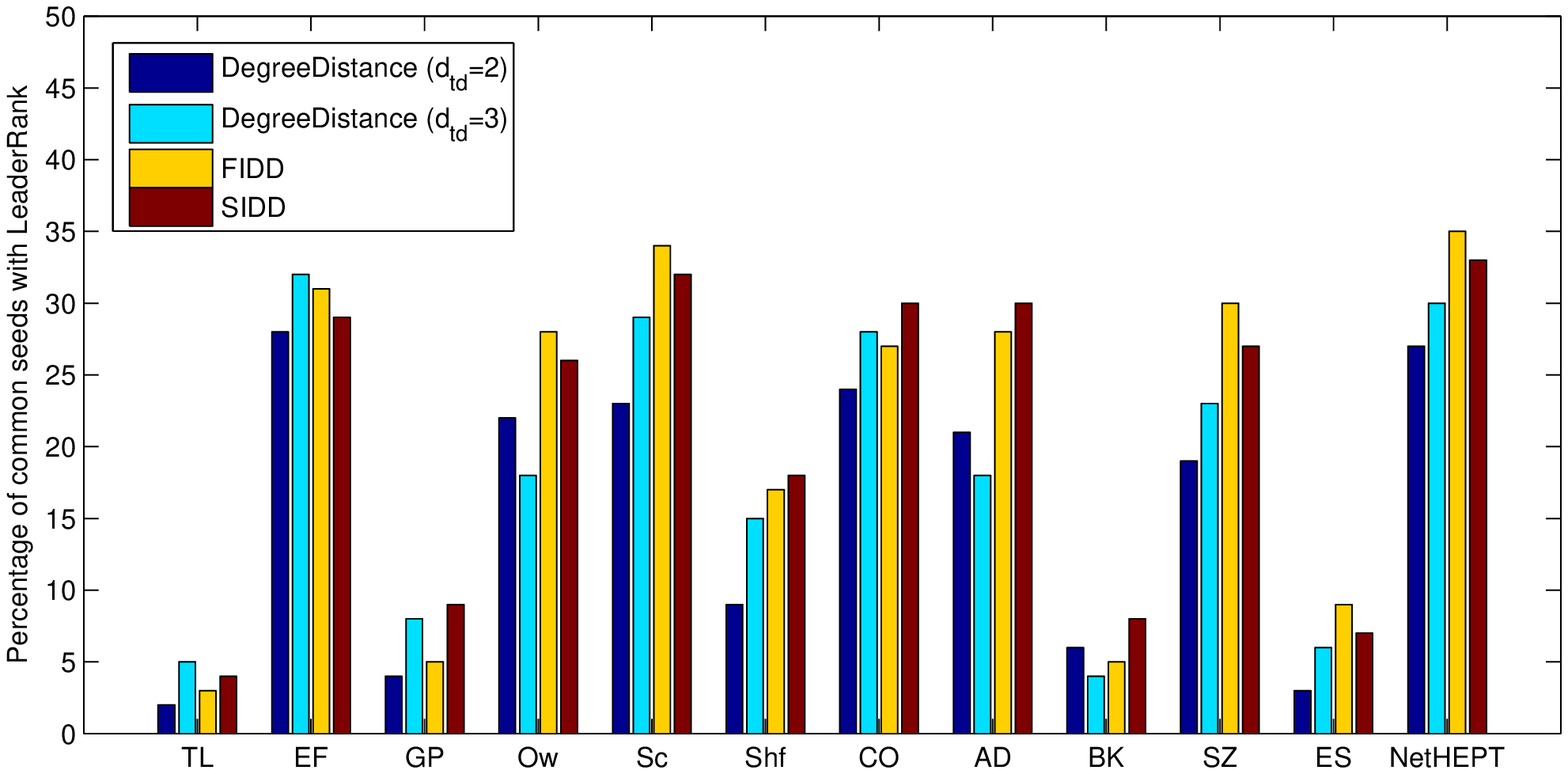}
              \caption{}
        \end{subfigure}
      &
      \begin{subfigure}[h]{2in}
              \centering
              \includegraphics[width=2in]{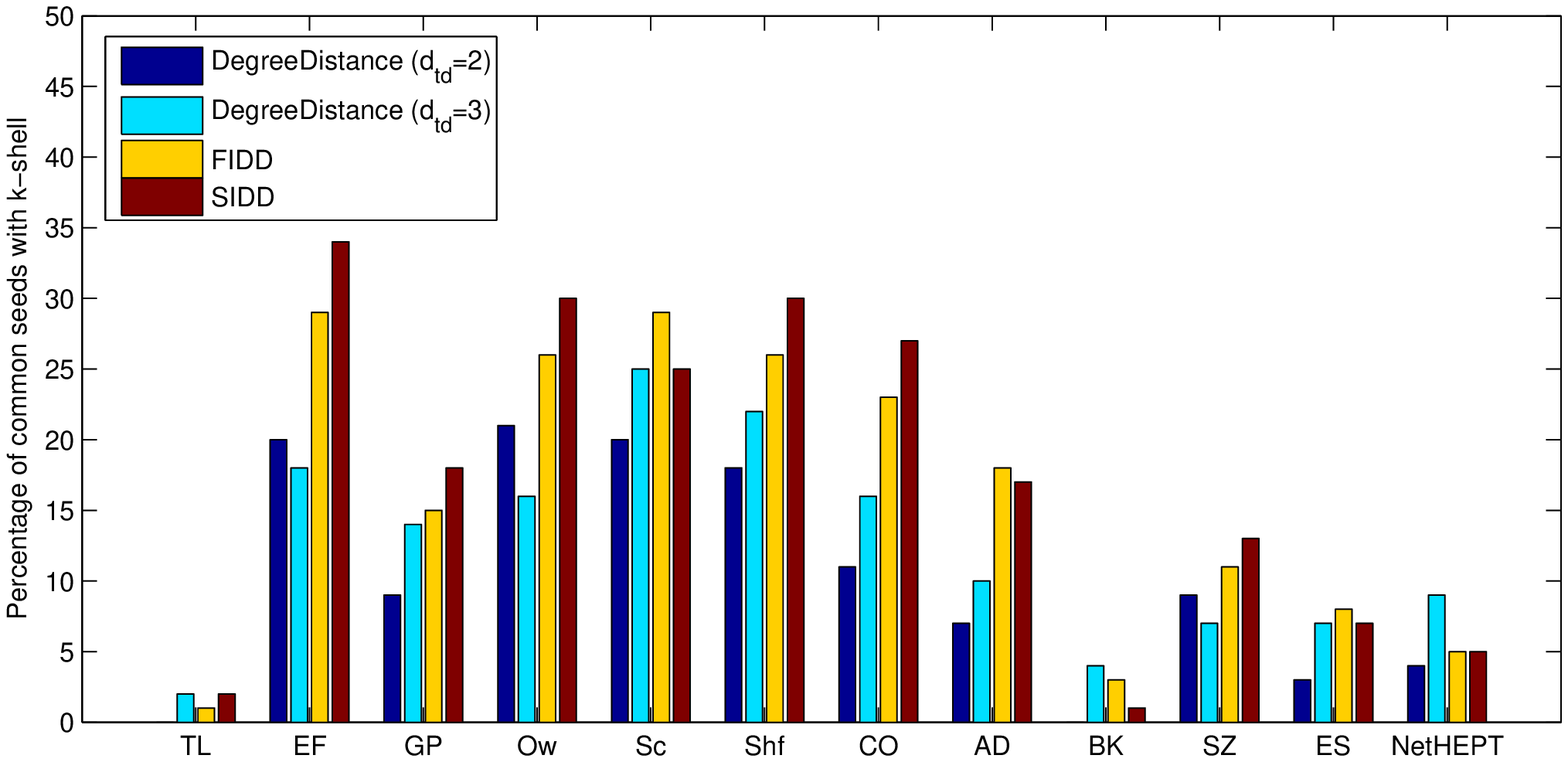}
              \caption{}
      \end{subfigure} 
  \end{tabular}
  \end{center}
  \caption{The number of common seeds of the seed sets of DegreeDistance, FIDD, SIDD, and the seed sets of (a) high-degree (b) betweenness (c) closeness (d) PageRank (e) LeaderRank (f) $k$-shell. Here the seed set size $k = 100$.} \label{commonseedsEvaluation}
  \end{figure}
  \end{center}

In Figure \ref{commonseedsEvaluation}, we have compared the number of common seeds between DegreeDistance, FIDD, SIDD seed sets and other measures' for $k = 100$. By looking back at Figure \ref{fig23}, we can see that the rate of having common seeds between our measures and other measures is looked up, and our methods choose almost different seeds. \\
The relationship between SIDD and some other measures is evaluated using the Pearson's correlation on three real-world datasets presented in Table \ref{tablePeC}.

  \begin{table}[H]
  \begin{center}
  \rowcolors{1}{white}{pink!30}
  \caption{The Pearson's correlation between SIDD and other measures on three datasets.}
   \label{tablePeC}
  \begin{tabular}{|lcccccc|}
  \hline
  \textsc{Dataset} & \textsc{Degree} & \textsc{Closeness} & \textsc{PageRank} & \textsc{DegreeDiscount} & \textsc{LeaderRank} &\textsc{$k$-shell} \\ \hline \hline
  BK & 0.431 & -0.343 & 0.35 & 0.53 & -0.009 & 0.27\\
  ES & 0.39 & -0.341 & 0.18 & 0.22 & -0.0421 & 0.31 \\
  SZ & 0.53 & 0.28 & 0.35 &  0.66 & 0.19 & 0.39 \\ \hline
  \end{tabular}
  \end{center}
  \end{table}

\subsection{Unique nodes influenced by DegreeDistance, FIDD, SIDD, and high-degree seeds}

\begin{figure}[H]
 \centering
 \includegraphics[width = 4in]{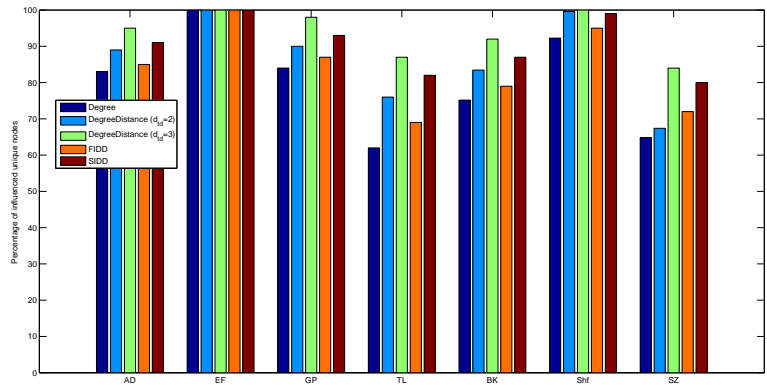}
 \caption{The percentage of unique nodes influenced by DegreeDistance seeds with $\DT \in \{ 2,3\}$, FIDD, SIDD, and high-degree seeds.} \label{fig:comp}
 \end{figure}

To evaluate DegreeDistance seeds in distance threshold $\DT = \{ 2,3 \}$ from each other, FIDD and SIDD seeds, we check the percentage of the unique nodes in the network that are influenced via them. From Figure \ref{fig:comp}, it is clear that in large-scale networks, DegreeDistance seeds with $\DT= 3$ and SIDD cover significantly more unique nodes in comparison with high-degree.

\subsection{Runtime performance and spread of influence by DegreeDistance, FIDD, and SIDD}

\begin{center}
  \begin{figure}[H]
  \begin{center}
  \begin{subfigure}[h]{3.2in}
  \centering
   \includegraphics[width = 3.2in]{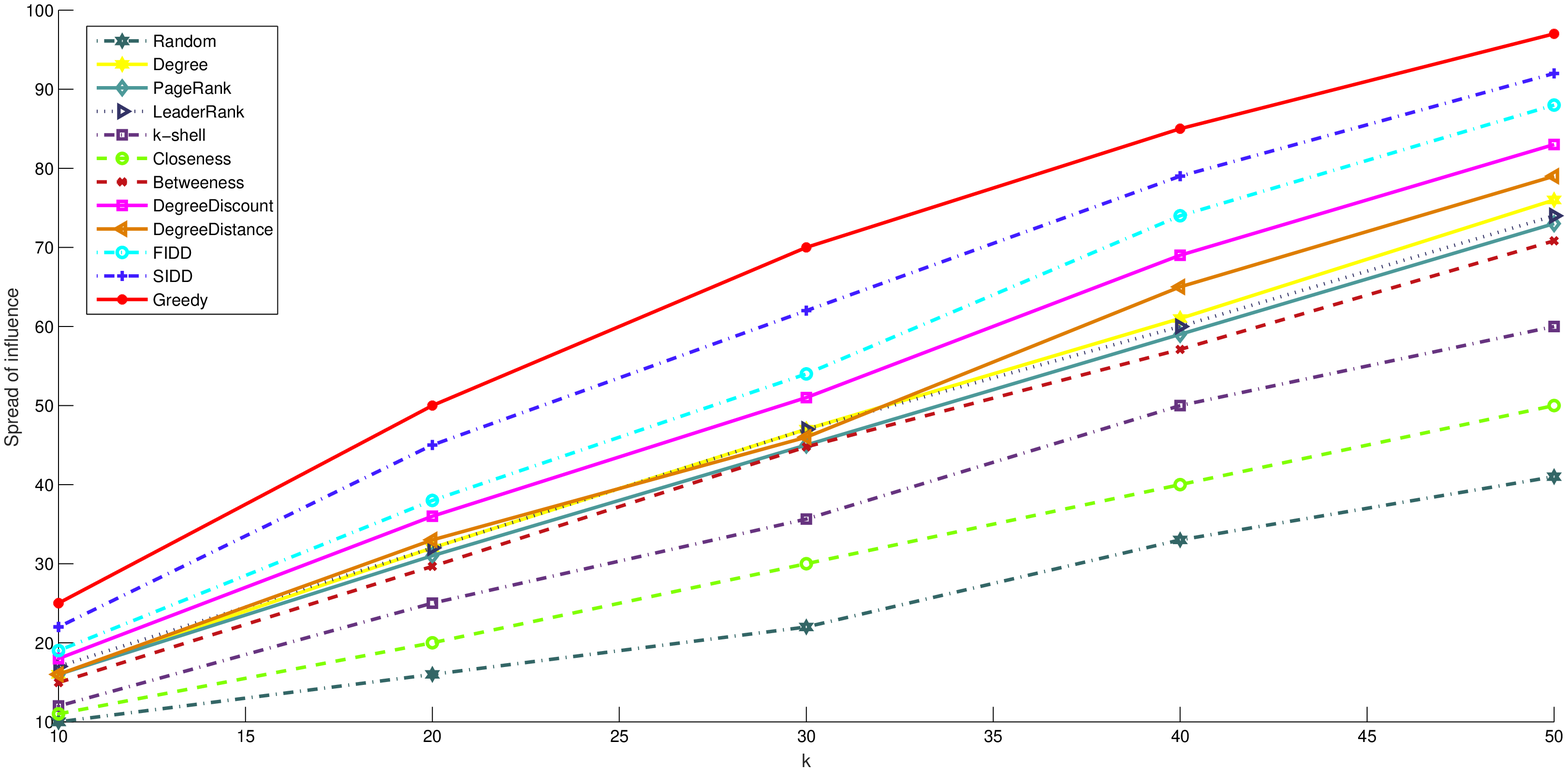}
   \caption{} \label{fig4}
  \end{subfigure}
  ~
   \begin{subfigure}[h]{3in}
    \centering
    \includegraphics[width = 3in]{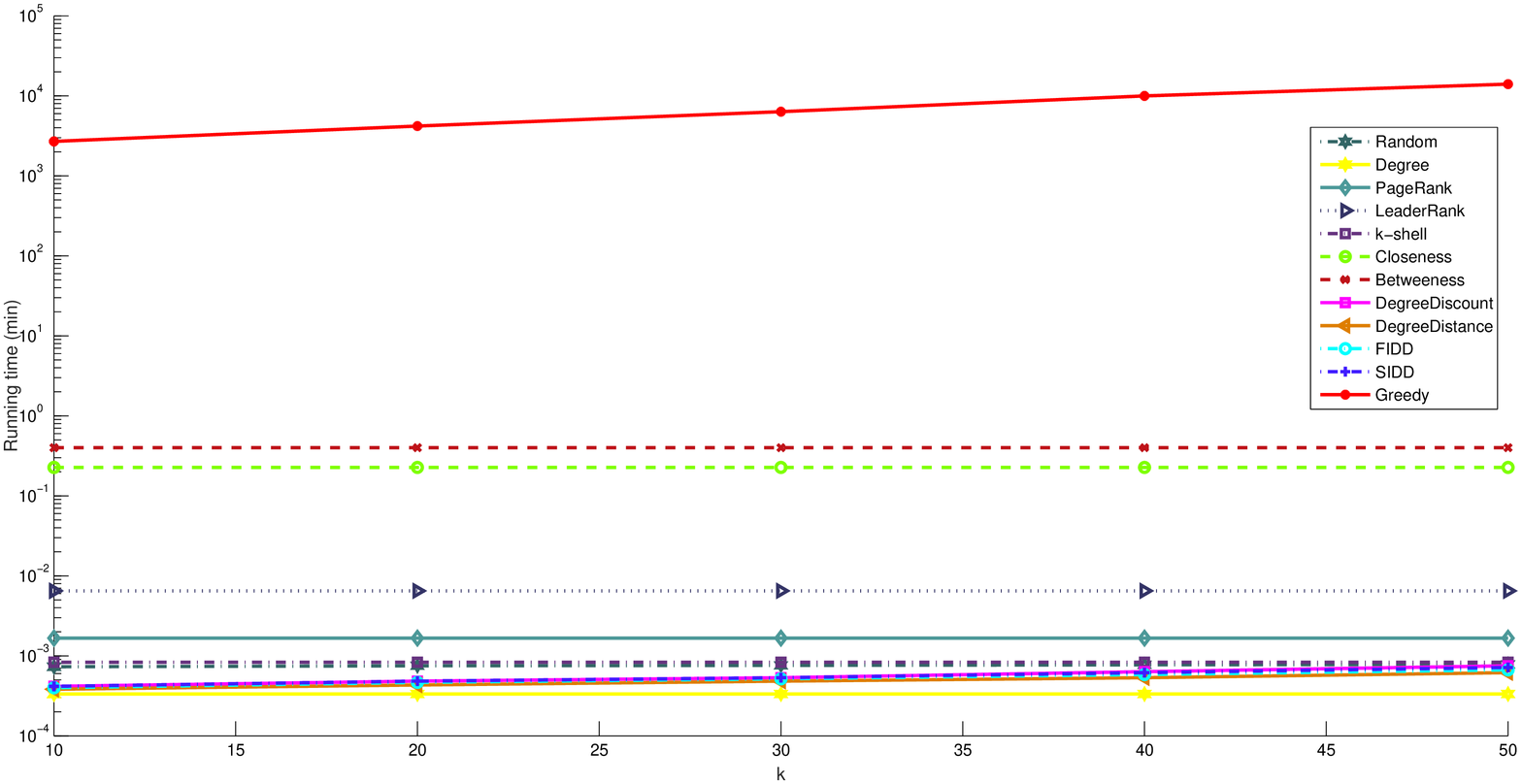}
     \caption{}\label{fig5}
    \end{subfigure}
    \end{center}
    \caption{(a) Comparison of spread of influence by seeds obtained from DegreeDistance, FIDD, and SIDD with other measures on NetHEPT dataset. (b) Comparison of runtime performance in order to identify seeds using DegreeDistance, FIDD, and SIDD with other centrality measures on the same dataset.} \label{fig46}
    \end{figure}
  \end{center}

\begin{center}
  \begin{figure}
  \begin{center}
  \begin{subfigure}[h]{3in}
  \centering
   \includegraphics[width = 3in]{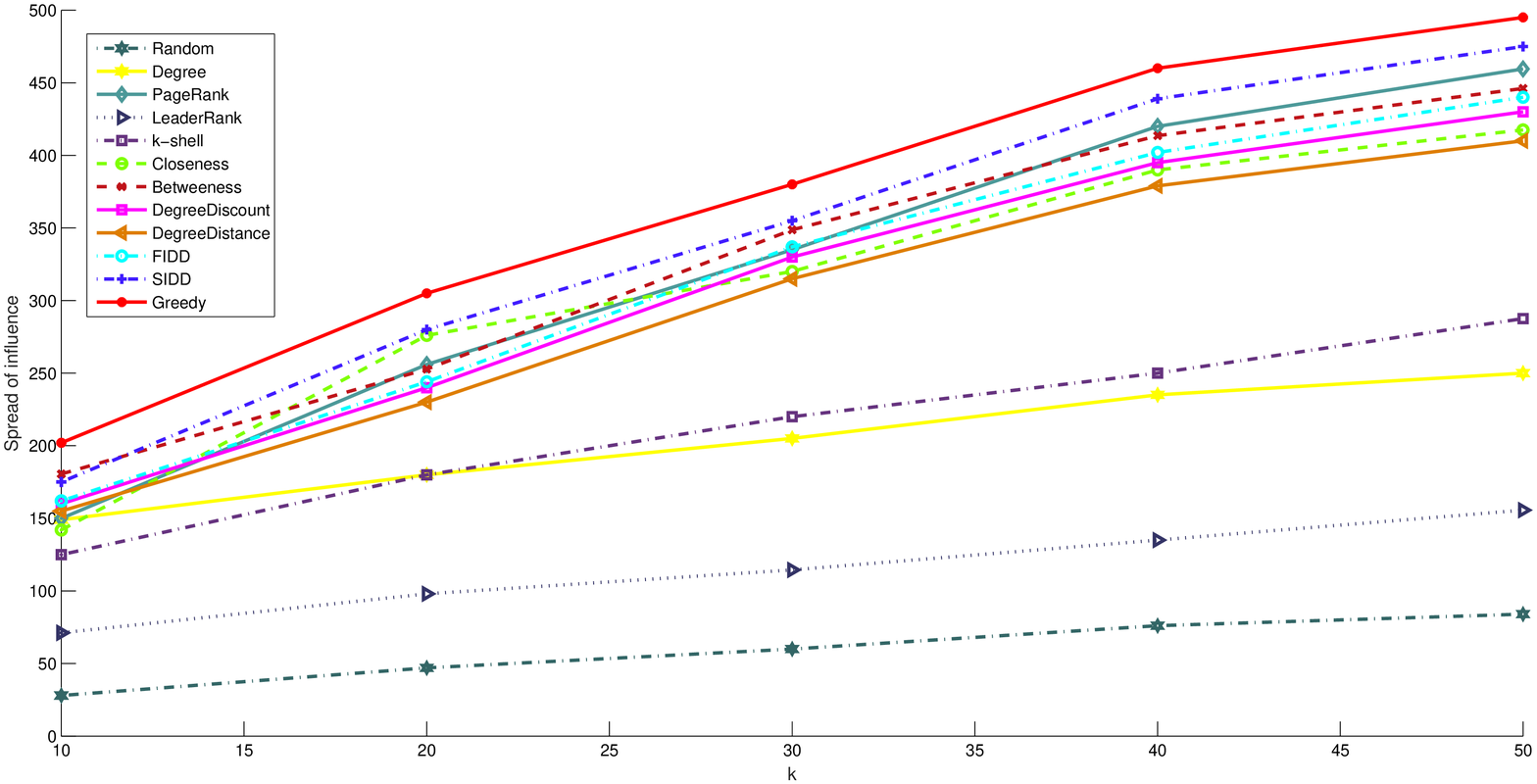}
  \caption{} \label{fig6}
  \end{subfigure}
  ~
  \begin{subfigure}[h]{3.1in}
  \centering
  \includegraphics[width = 3.1in]{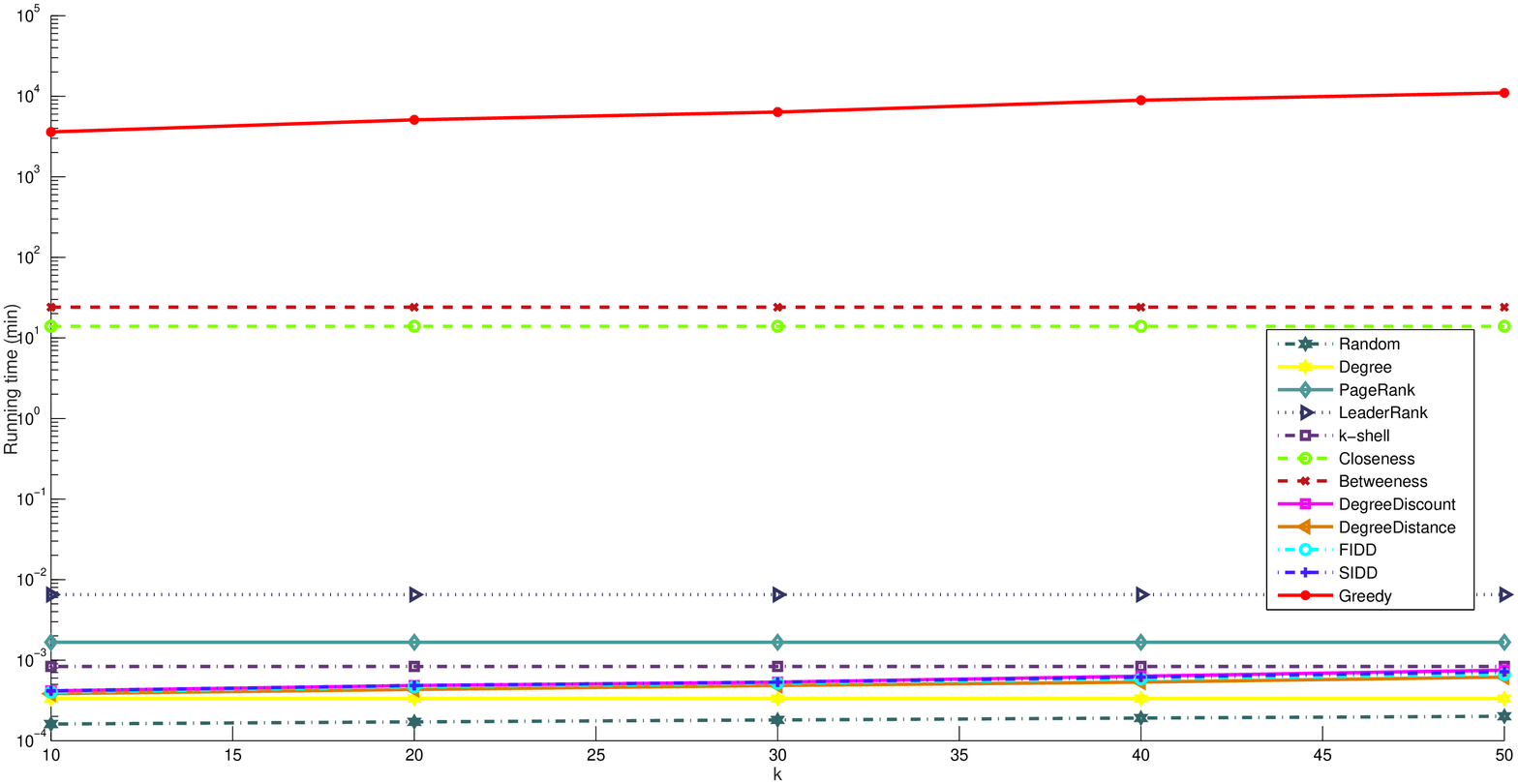}
  \caption{} \label{fig7}
  \end{subfigure}
  \end{center}
  \caption{(a) Comparison of spread of influence by seeds obtained from DegreeDistance, FIDD, and SIDD with other measures on BK dataset. (b) Comparison of runtime performance in order to identify seeds using DegreeDistance and its improvements with other measures on the same dataset. } \label{fig67}
  \end{figure}
  \end{center}

To evaluate the spread ability of DegreeDistance, FIDD, and SIDD, we compare them not only with other well-known measures, but with random method ($k$ random nodes form the seed set)
under the independent cascade (IC) model \cite{kempe2003maximizing} to simulate the influence propagation with a 10'000-iteration process for each seed set and take the average of all the influence spreads. To analyze the spread efficiency of the mentioned methods, which are depicted in Figures \ref{fig46}, \ref{fig67}, and Table \ref{tablespreading}, we apply them to some large-scale datasets from Table \ref{table1}. The value of $\theta$ is assumed to be equal to the average degree of the network. Figures \ref{fig46} and \ref{fig67} show the spread effectiveness and runtime efficiency on NetHEPT and BK datasets, respectively.

In our experiments, the influence score of a seed, $v$, on each $w \in \Ne^{(1)} (v)$ is set to be the fixed value $0.01$, that is Eq. (\ref{eq5}) becomes 
$$
\textnormal{inf} =
\begin{cases}
 0.01 + \Big( (0.01)^2 \cdot | \textnormal{CN}^{(1)}(s^\prime, v)| \Big), & \textnormal{ if $v$ and $s^\prime$ are adjacent,} \\
  (0.01)^2 \cdot | \textnormal{CN}^{(1)}(s^\prime, v)|, & \textnormal{ otherwise.} \\
\end{cases}
$$

 From this figure, one can find out that in spite of random model which has the lowest spread ability of influence, greedy method has the highest propensity. Clearly, greedy method is exceedingly time-consuming and is not an appropriate measure for large-scale networks. Therefore, we have not taken these two measures any farther. In addition to the high speed performance of DegreeDistance (especially SIDD), it has a satisfactorily close spread ability of influence compared to greedy method. The running time of each algorithm is illustrated in Fig \ref{fig46}. The experiments are carried out on a state-of-the-art desktop machine with Intel Core i7 3.4 GHz CPU and 4GB RAM.

  From the above arguments, we conclude that our proposed centrality measure and its improvements have a satisfying, acceptable performance in comparison to other methods.

  As we mentioned, the influence score is set as $0.01$. By increasing this value, we can avoid selecting seed nodes that have a high influence on one another and consequently we observe more difference between FIDD and SIDD. \\

\begin{center}
  \begin{figure}
  \begin{center}
  \begin{tabular}{ccc}
  \begin{tabular}{c}
      \begin{subfigure}[h]{2in}
      \centering
      \includegraphics[width=2in]{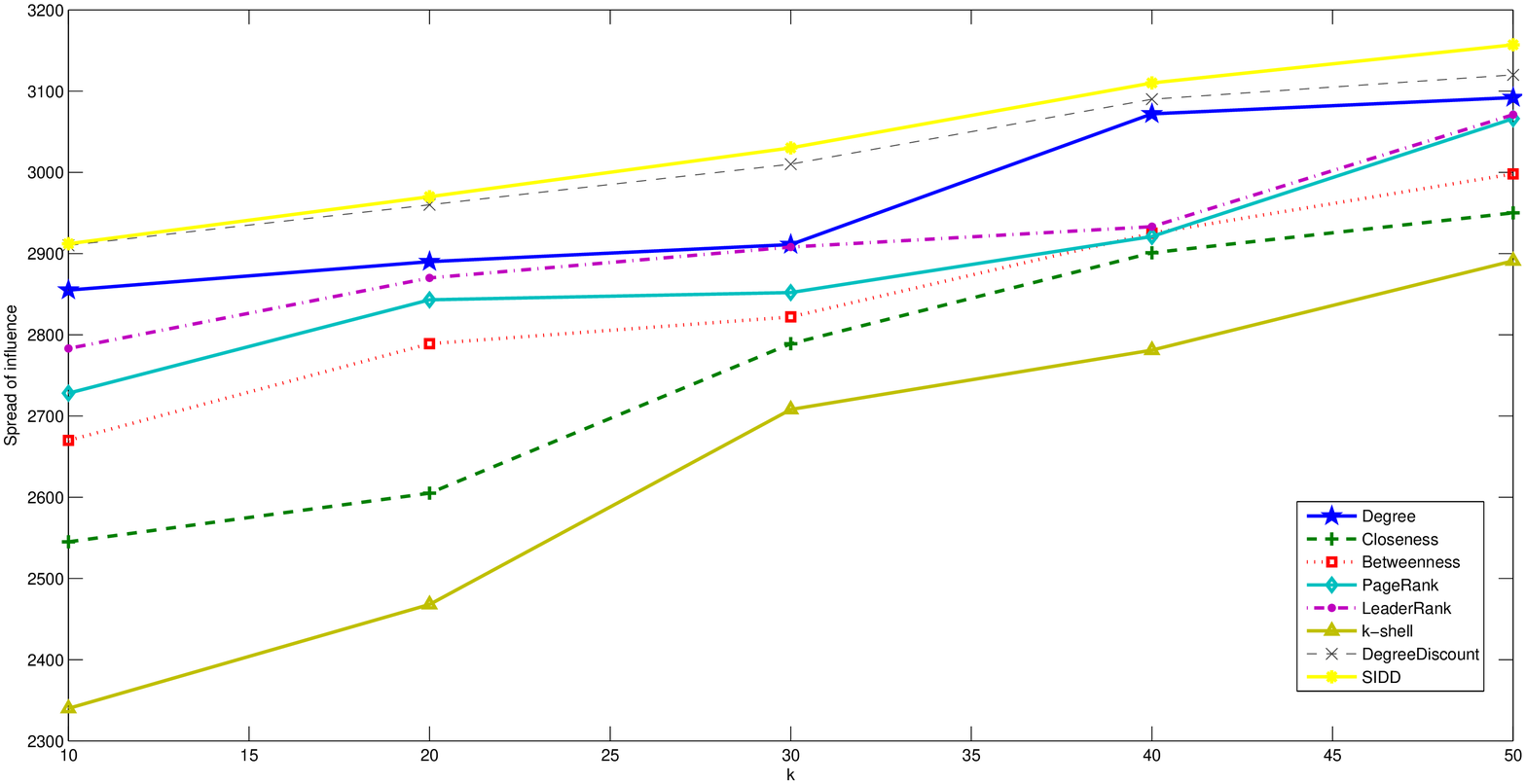}
      \end{subfigure}
	\\
      \begin{subfigure}[h]{2in}
      \centering
      \includegraphics[width=2in]{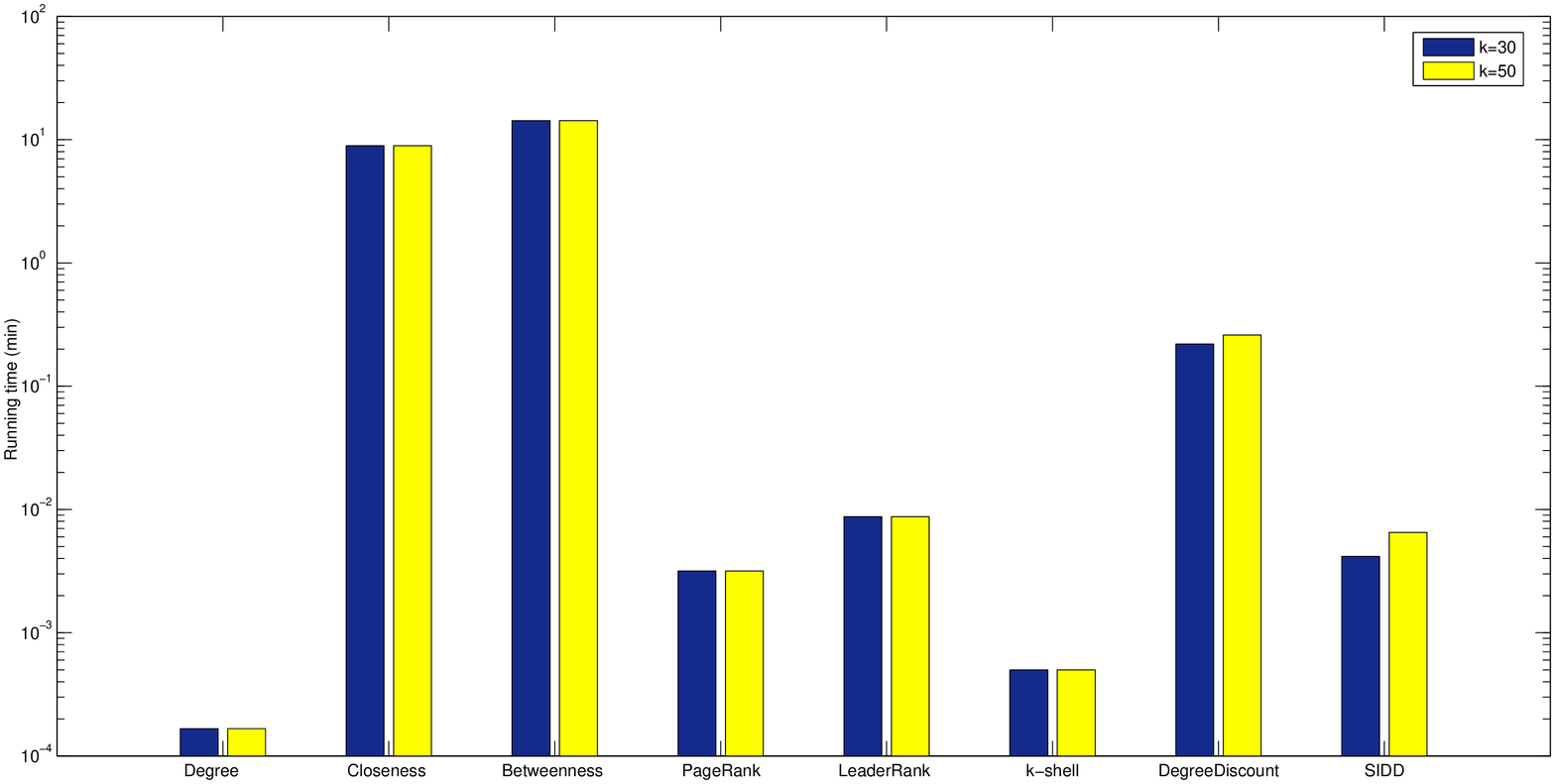}
      \caption{ES}
      \end{subfigure}
   \end{tabular}
   &
         \begin{tabular}{c}
               \begin{subfigure}[h]{2in}
                   \centering
                   \includegraphics[width=2in]{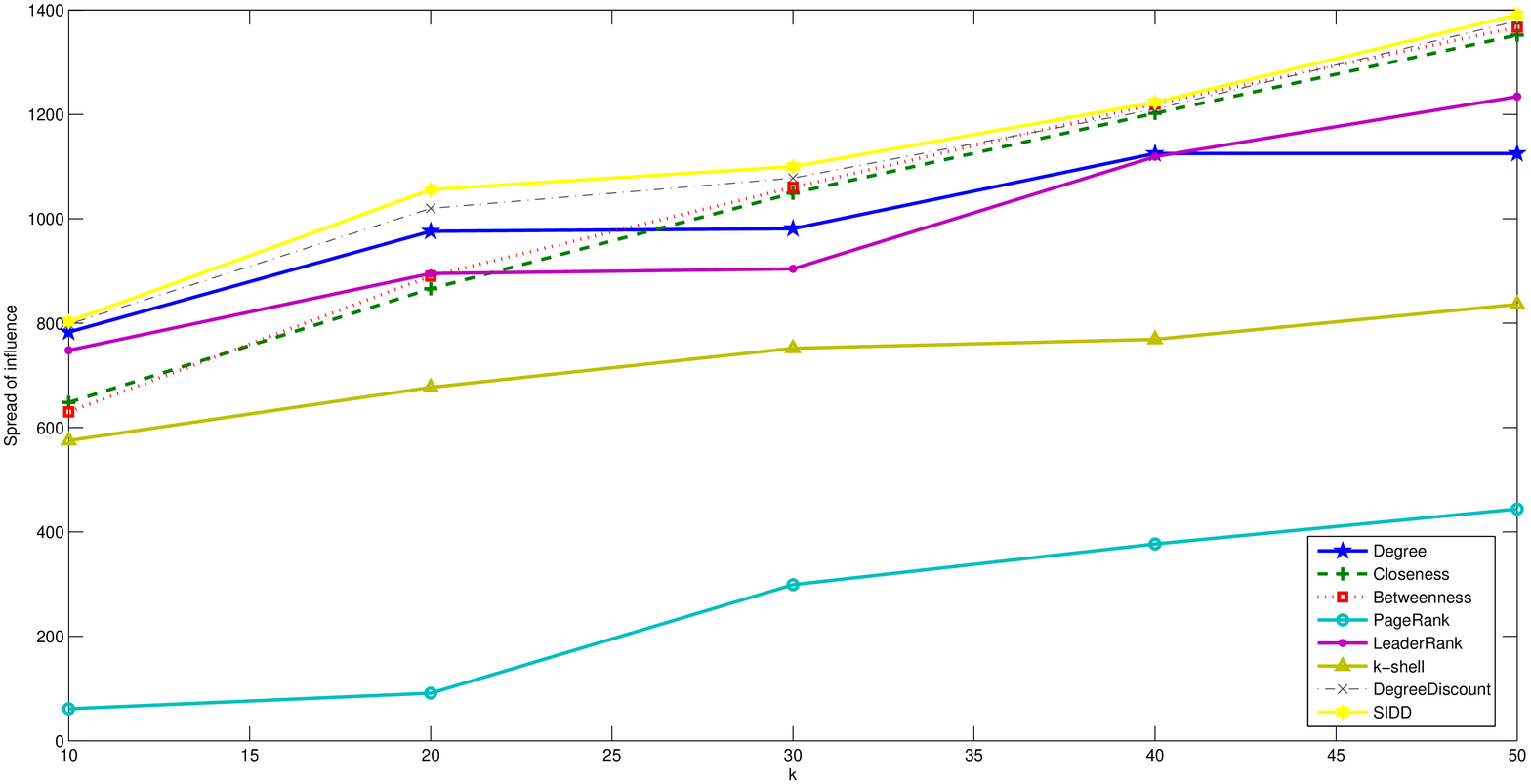}
                \end{subfigure}
            \\
               \begin{subfigure}[h]{2in}
                   \centering
               \includegraphics[width=2in]{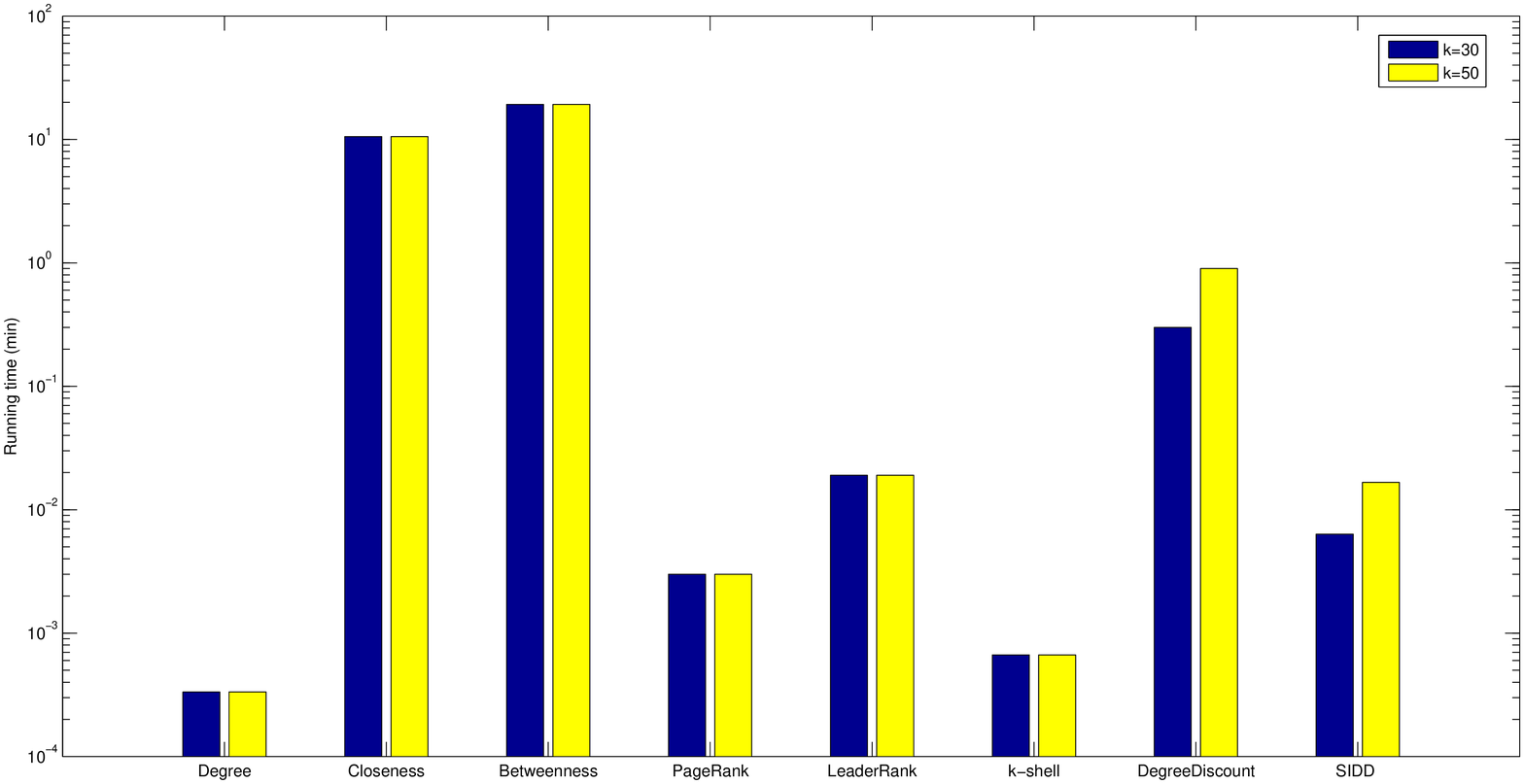}
                \caption{SZ}
               \end{subfigure}
        \end{tabular}
          &
           \begin{tabular}{c}	
                    \begin{subfigure}[h]{2in}
                           \centering
                           \includegraphics[width=2in]{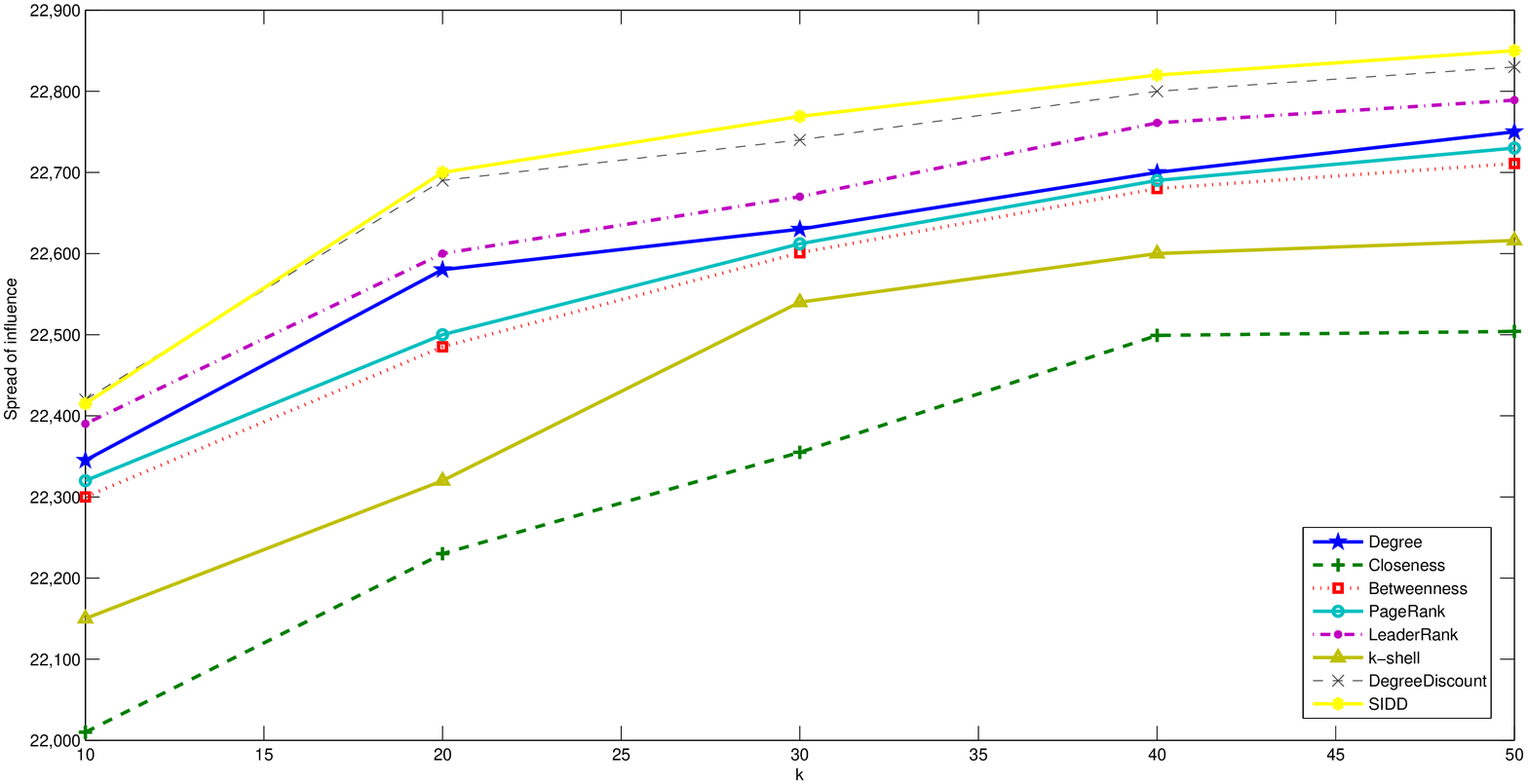}
                     \end{subfigure}
                   \\
                   \begin{subfigure}[h]{2in}
                           \centering
                           \includegraphics[width=2in]{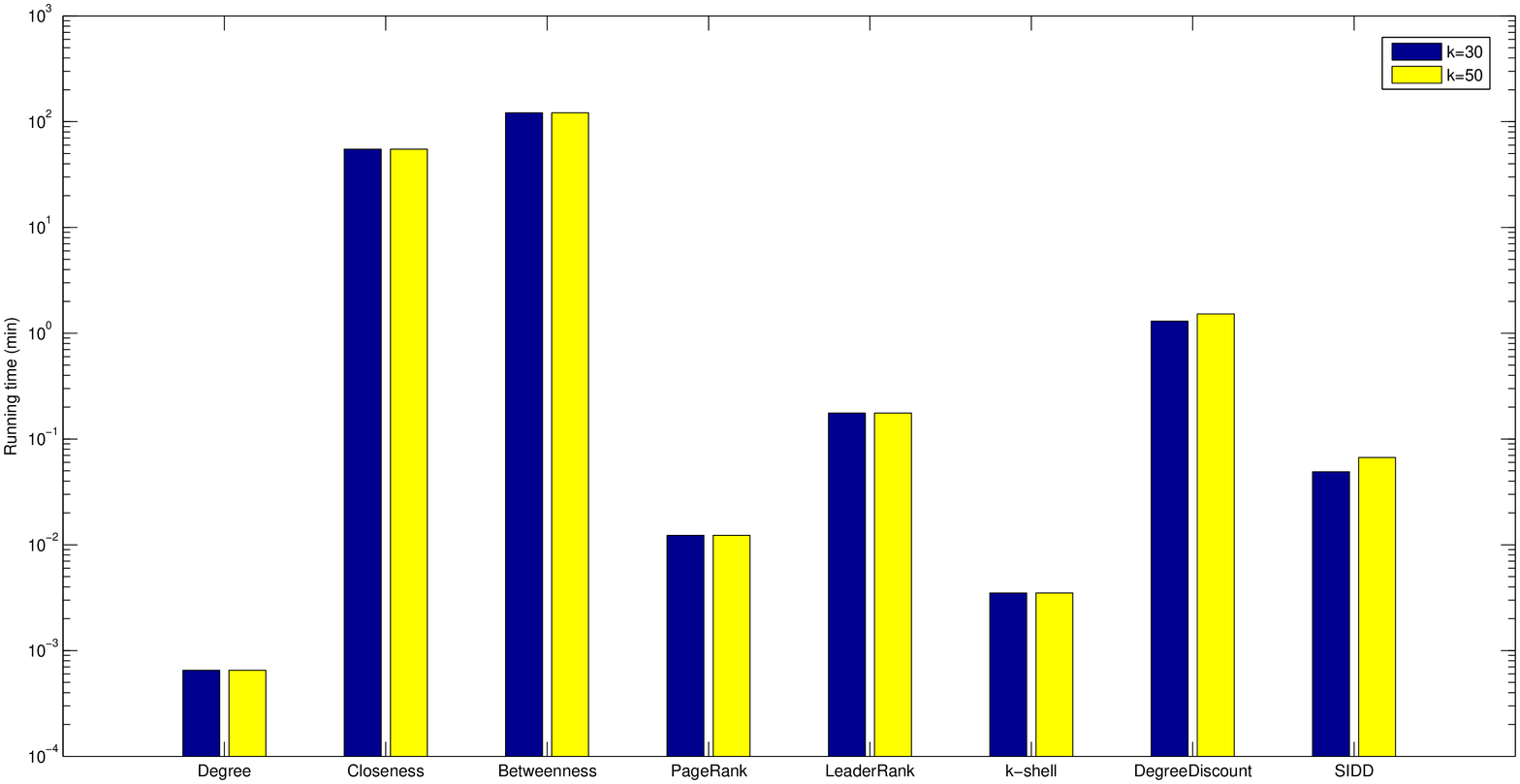}
                   \caption{CO}
                   \end{subfigure} 
         \end{tabular} 
   \\
	\begin{tabular}{c}	
       \begin{subfigure}[h]{2in}
              \centering
              \includegraphics[width=2in]{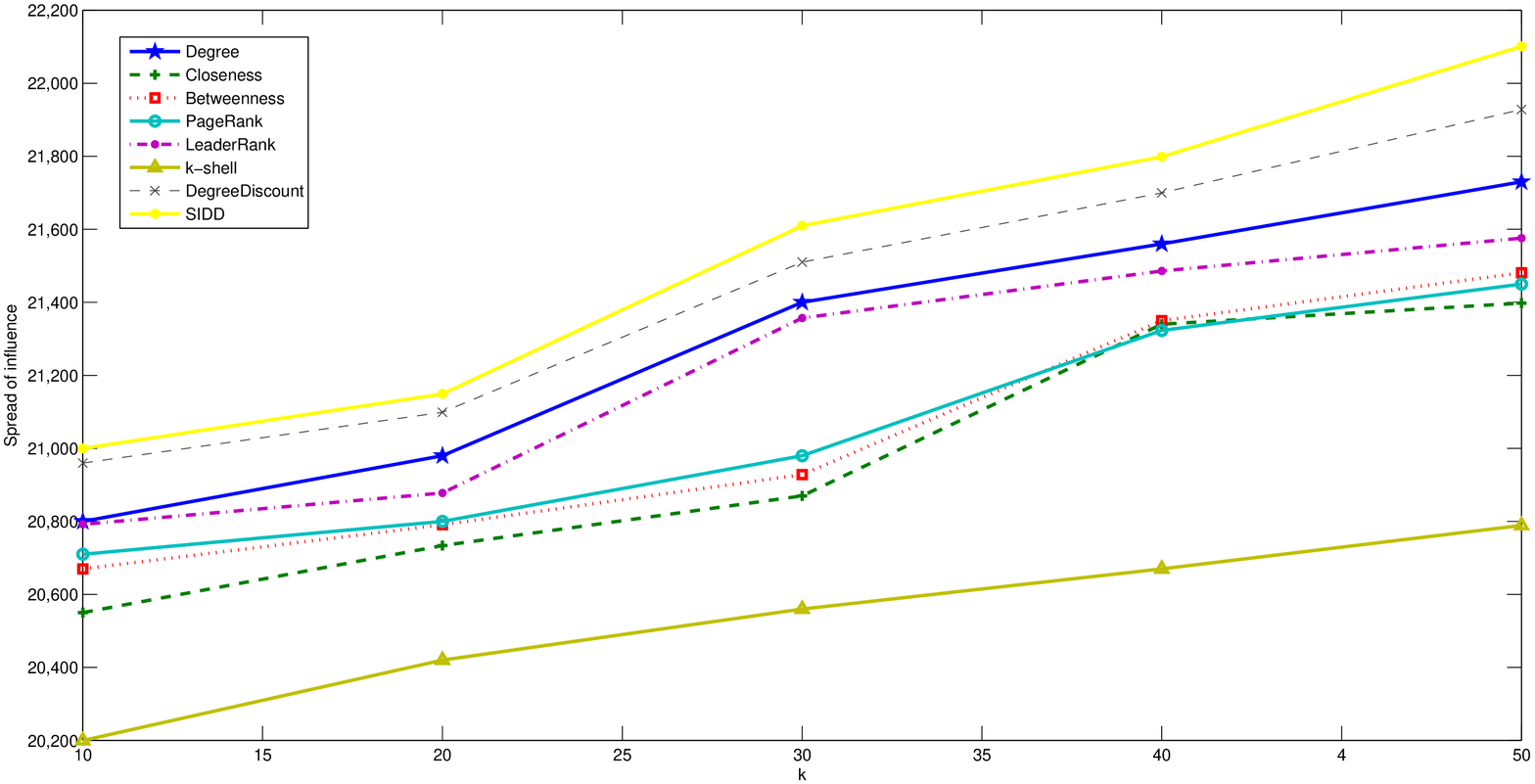}
        \end{subfigure}
      \\
      \begin{subfigure}[h]{2in}
              \centering
              \includegraphics[width=2in]{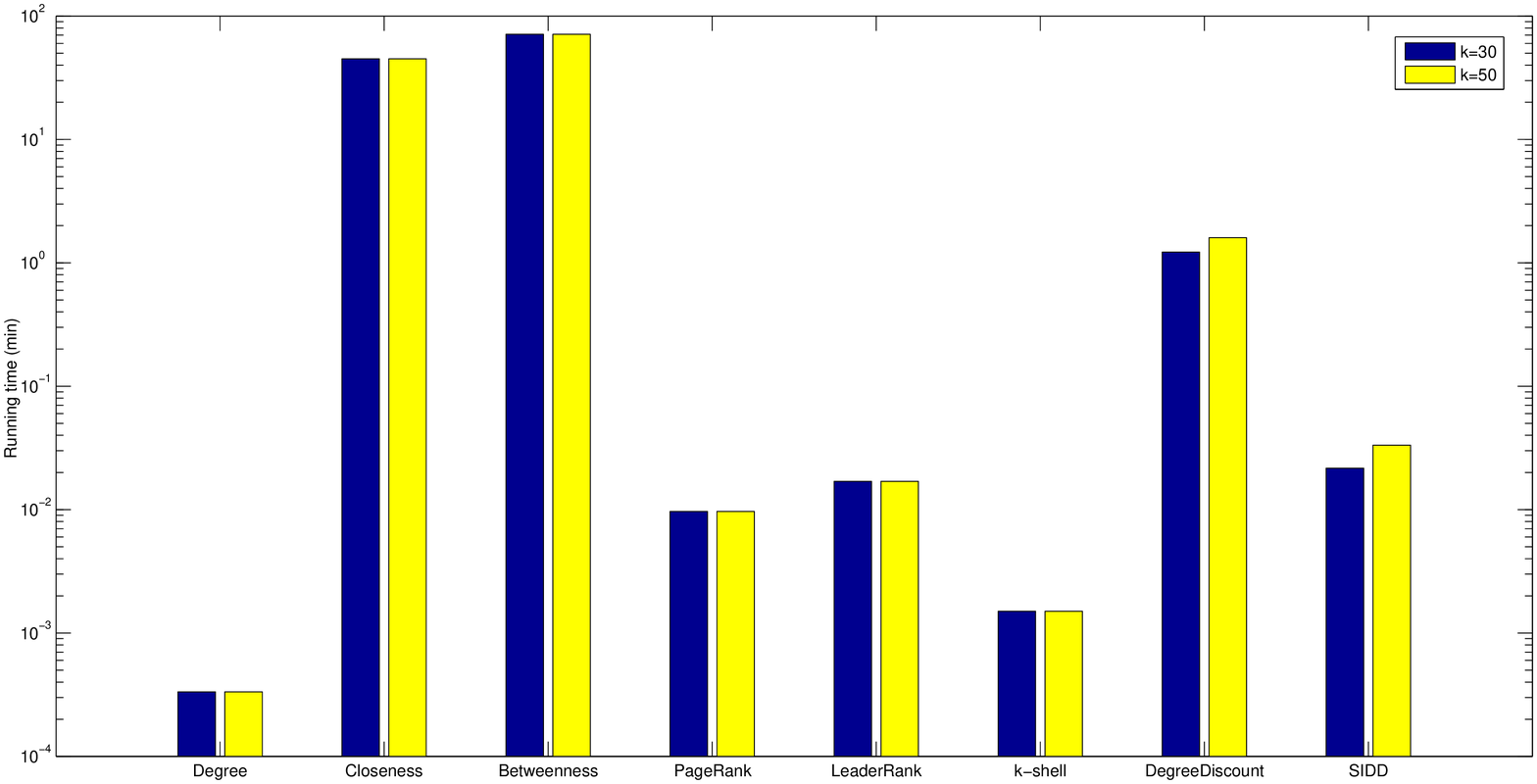}
      \caption{Fl}
      \end{subfigure} 
     \end{tabular}
 &
      \begin{tabular}{c}
      \begin{subfigure}[h]{2in}
                  \centering
                  \includegraphics[width=2in]{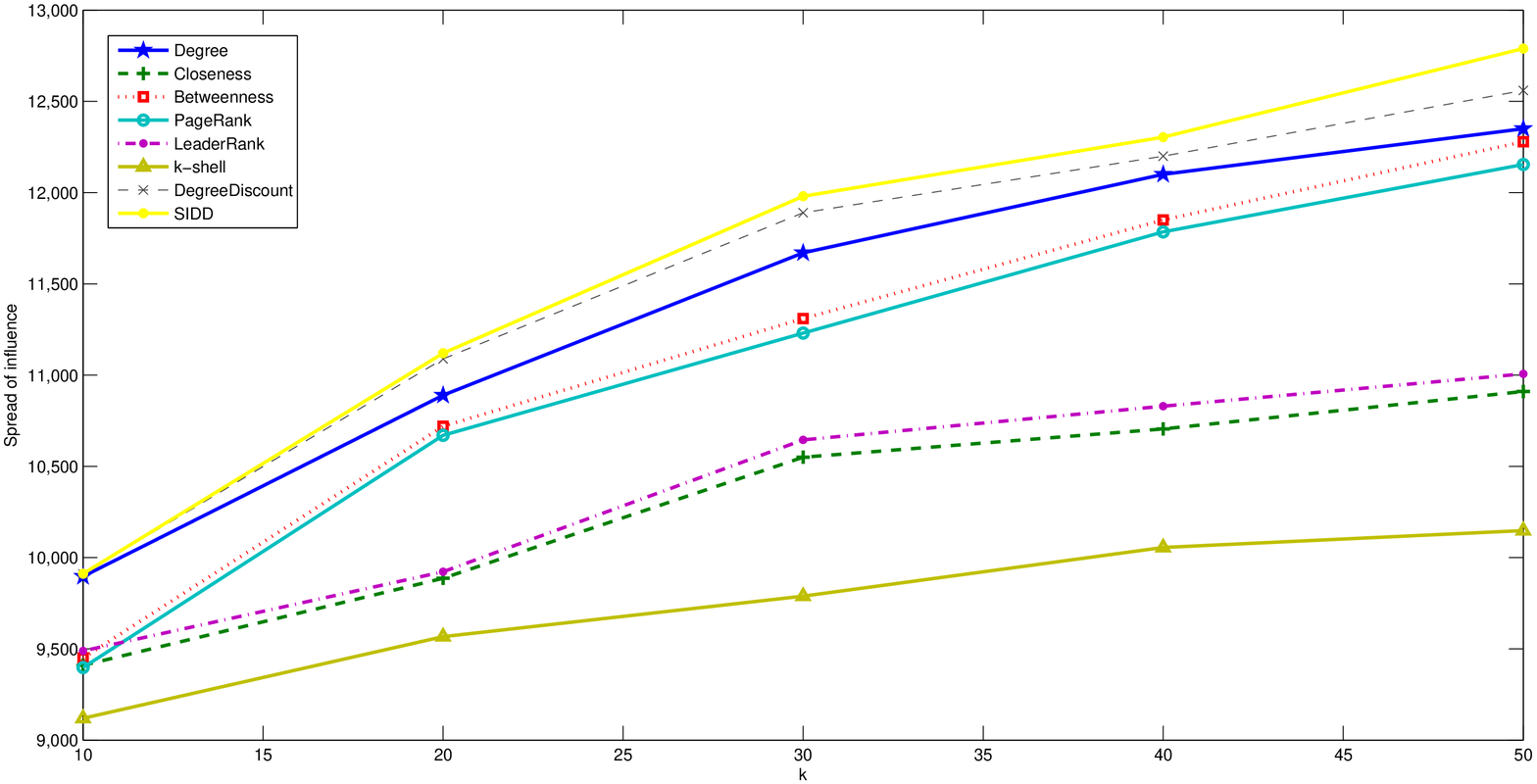}
        \end{subfigure}
        \\
       \begin{subfigure}[h]{2in}
          \centering
      \includegraphics[width=2in]{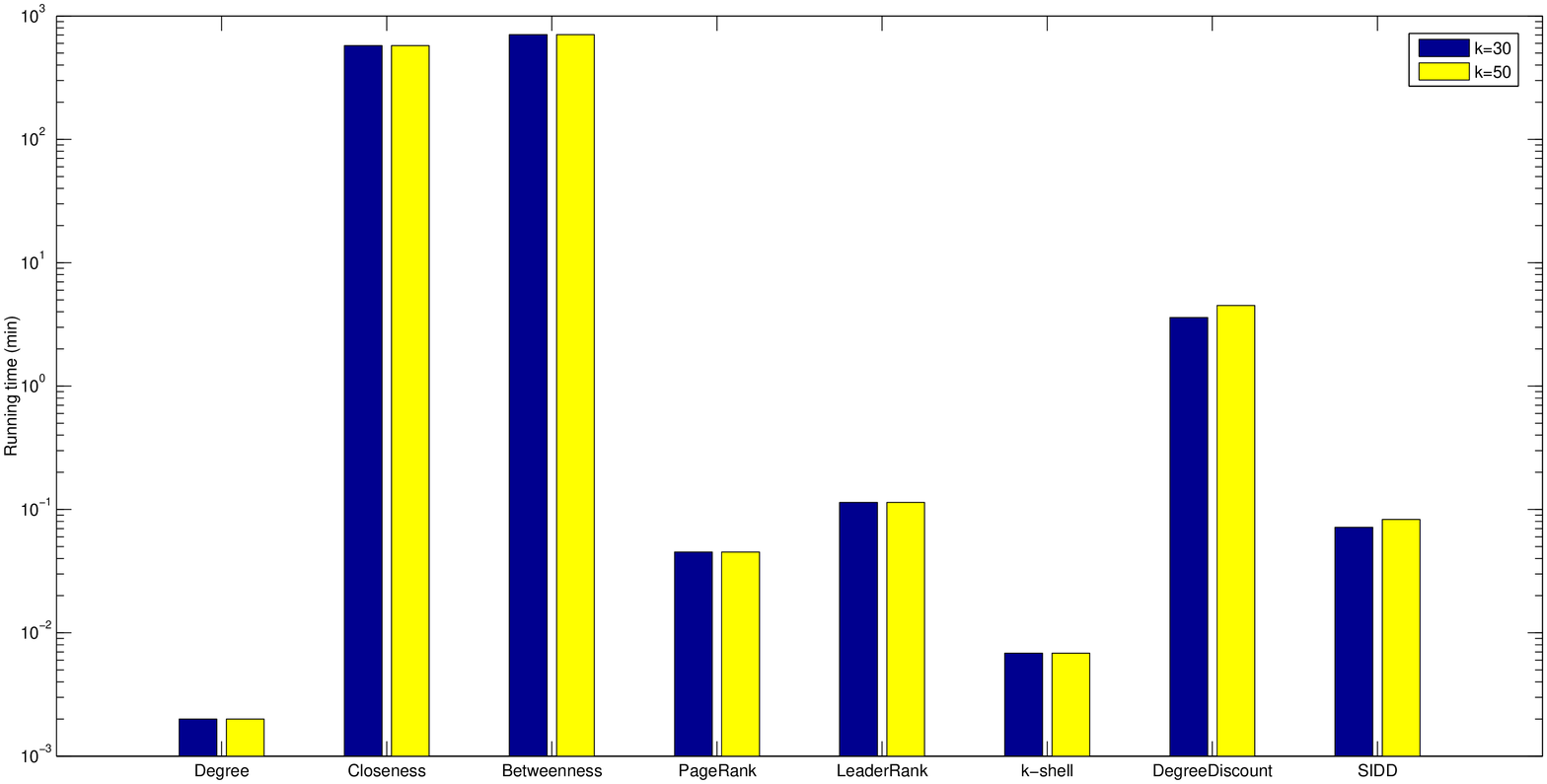}
      \caption{CY}
      \end{subfigure}
      \end{tabular}
      &
	\begin{tabular}{c}
      \begin{subfigure}[h]{2in}
            \centering
            \includegraphics[width=2in]{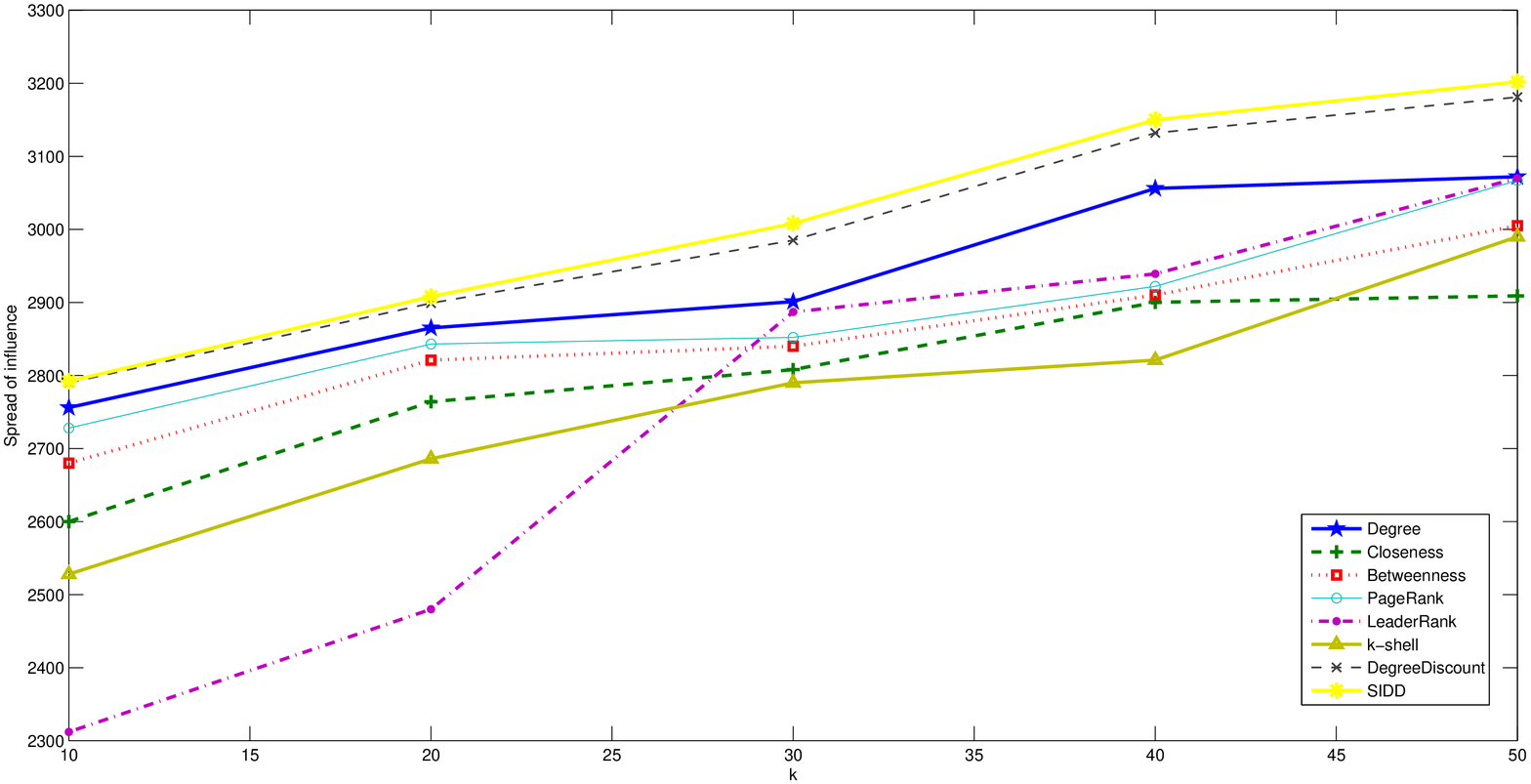}
      \end{subfigure}
      \\
      \begin{subfigure}[h]{2in}
        \centering
        \includegraphics[width=2in]{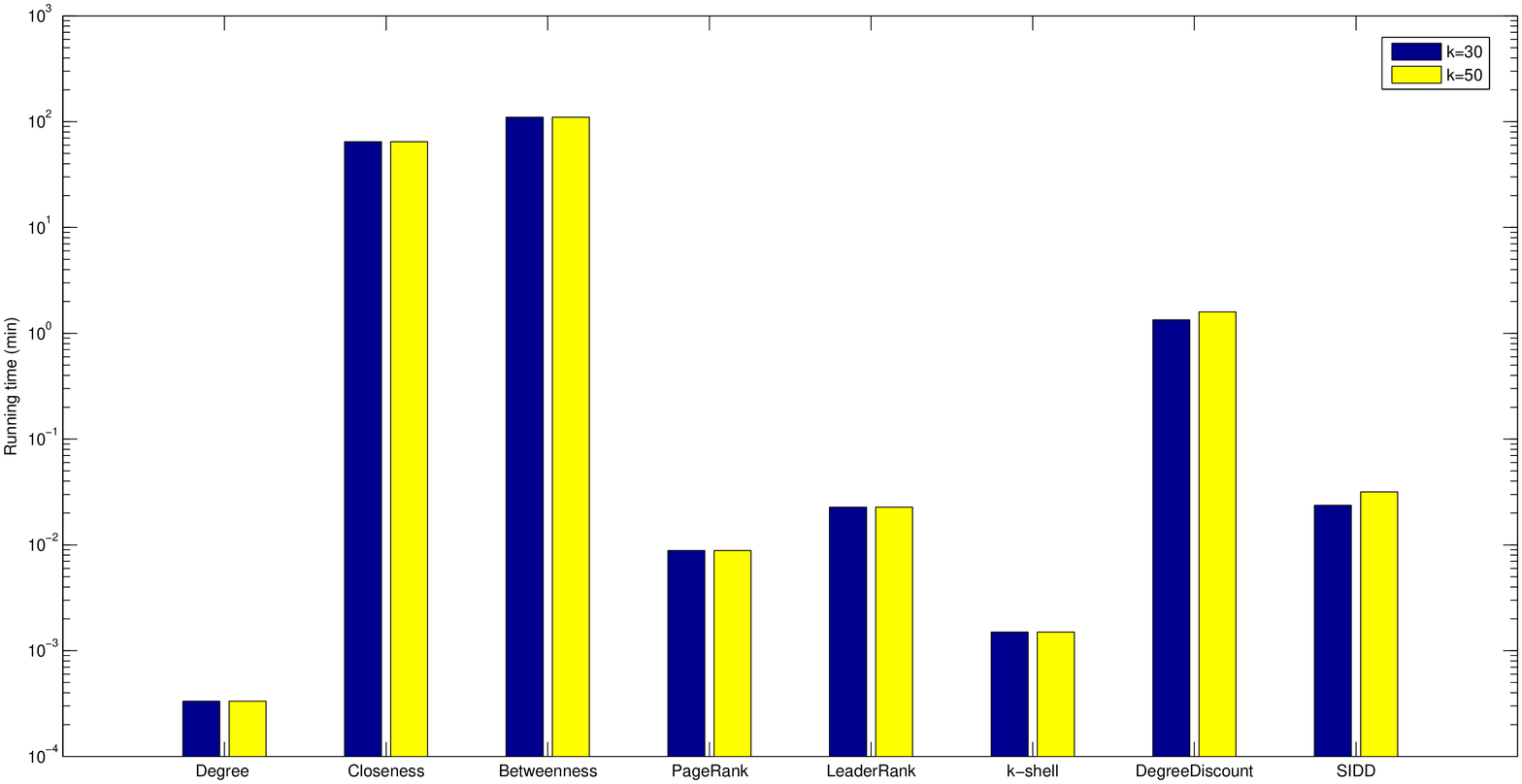}
        \caption{GW}
      \end{subfigure}
      \end{tabular}
  \end{tabular}
  \end{center}
  \caption{SIDD outperforms other measures with respect to maximizing spread of influence, which demonstrates a more precise selection of seeds. Its running time is quite legitimate for $k = 30, 50$.} \label{figlargedatasets}
  \end{figure}
  \end{center}

  Based on the above discussion and evaluation of our method on the two datasets, NetHEPT and BK, SIDD outperforms the previous two versions. Thereof, for further evaluation, we have considered SIDD only and carried out the same experiments as the previous two ones on more datasets (see Figure \ref{figlargedatasets}. Here the running time is computed for $k=30,50$) as well as the numerical comparison of the spread effectiveness of SIDD with other methods on the same datasets (Table \ref{tablespreading}).

  The results assert the superiority of SIDD over the other methods. Due to the close distance of nodes in the seed sets obtained from other measures, by increasing the size of the dataset, we do not see much spreading progress; for example, by applying the closeness measure on Gowalla, when we change $k=40$ to $k=50$, only nine more nodes got activated, or similarly $k$-shell decomposition does not show satisfactory promotion, the reason is that it gives the key users topologically in the inner-core of the network. Although the seed nodes (with high $k$-shell index) have high spread ability individually, we observe that these nodes are mostly in close neighborhood of one another, and they hence all together (top $k$) do not display a good spreading effectiveness compared to other commonly used measures of influence.

\begin{landscape}
\begin{table}[H]
\begin{center}
\caption{Spreading effectiveness of seeds by different seed sets.}
 \label{tablespreading}
\begin{tabular}{|lcccccccc>{\columncolor{pink!30}}c|}
\hline
\textsc{Dataset} & \textsc{Top $k$} & \textsc{Degree} &  \textsc{Closeness} &  \textsc{Betweenness} & \textsc{DegreeDiscount} & \textsc{PageRank} & \textsc{LeaderRank} & \textsc{$k$-shell} & \textsc{SIDD} \\ \hline\hline
\multirow{5}{*}{ES} & $k=10$ & 2855 & 2545 & 2670 & 2910 & 2728 & 2783 & 2340 & 2912 \\
  & $k=20$ & 2890 & 2605 & 2789 & 2960 & 2843 & 2870 & 2468 & 2970  \\
  & $k=30$ & 2911 & 2789 & 2822 & 3010 & 2852 & 2908 & 2708 & 3030 \\
  & $k=40$ & 3072 & 2901 & 2925 & 3090 & 2921 & 2933 & 2781 & 3110  \\
  & $k=50$ & 3092 & 2950 & 2998 & 3120 & 3066 & 3071 & 2891 & 3157  \\ \hline
  \multirow{5}{*}{SZ} & $k=10$  & 783 & 648 & 630 & 798 & 61 & 748 & 575 & 803 \\
  & $k=20$ & 976 & 866 & 890 & 1020 & 91 & 895 & 677 & 1056  \\
  & $k=30$ & 981 & 1049 & 1060 & 1078 & 299 & 904 & 752 & 1100  \\
  & $k=40$ & 1125 & 1202 & 1220 & 1210 & 377 & 1119 & 769 & 1223 \\
  & $k=50$ & 1125 & 1352 & 1367 & 1379 & 444 & 1234 & 836 & 1391  \\ \hline
  \multirow{5}{*}{CO} & $k=10$  & 23865 & 23521 & 23814 & 24031 & 23836 & 23920 & 23661 & 24091 \\
    & $k=20$ &  24100 & 23741 & 23999 & 24301 & 24016 & 24130 & 23831 & 24371 \\
    & $k=30$ & 24150 & 23866 & 24115 & 24351 & 24128 & 24200 & 24031 & 24440  \\
    & $k=40$ & 24220 & 24010 & 24194 & 24411 & 24206 & 24291 & 24111 & 24491\\
    & $k=50$ & 24270 & 24015 & 24225 & 24441 & 24236 & 24319 & 24127 & 24541  \\ \hline
   \multirow{5}{*}{Fl} & $k=10$ & 20800 & 20550 & 20670 & 20960 & 20710 & 20792 & 20200 & 21000 \\
    & $k=20$ & 20980 & 20734 & 20791 & 21099 & 20800 & 20878 & 20420 & 21149  \\
    & $k=30$ & 21400 & 20870 & 20928 & 21510 & 20980 & 21357 & 20560 & 21610  \\
    & $k=40$ & 21560 & 21340 & 21350 & 21700 & 21323 & 21486 & 20670 & 21799  \\
    & $k=50$ & 21730 & 21398 & 21481 & 21928 & 21450 & 21576 & 20789 & 22101  \\ \hline
\multirow{5}{*}{CY} & $k=10$ & 9897 & 9410 & 9450 & 9910 & 9398 & 9489 & 9120 & 9912 \\
  & $k=20$ &  10890 & 9887 & 10720 & 11089 & 10670 & 9923 & 9567 & 11120  \\
  & $k=30$ & 11670 & 10550 & 11310 & 11890 & 11230 & 10645 & 9789 & 11980  \\
  & $k=40$ & 12100 & 10705 & 11850 & 12200 & 11785 & 10830 & 10056 & 12304  \\
  & $k=50$ & 12350 & 10910 & 12279 & 12560 & 12154 & 11007 & 10148 & 12789 \\ \hline
\multirow{5}{*}{GW} & $k=10$ & 2756 & 2600 & 2680 & 2790 & 2728 & 2312 & 2528 & 2792 \\
  & $k=20$ & 2865 & 2764 & 2821 & 2899 & 2843 & 2480 & 2686 & 2908  \\
  & $k=30$ & 2911 & 2808 & 2840 & 2985 & 2852 & 2887 & 2790 & 3008 \\
  & $k=40$ & 3072 & 2900 & 2910 & 3132 & 2922 & 2939 & 2821 & 3150 \\
  & $k=50$ & 3092 & 2909 & 3005 & 3181 & 3067 & 3070 & 2990 & 3202 \\ \hline
\end{tabular}
\end{center}
\end{table}
\end{landscape}

\section{Conclusions and future directions}
\label{conclusion}

In this paper, we presented an overviewed some well-known measures such as high-degree, betweenness, closeness, eigenvector, PageRank, DegreeDiscount, LeaderRank, and $k$-shell. Using ten datasets, we verified that the seed sets obtained by these measures have many seeds in common. We also showed that in the seed sets, the cardinalities $|\textnormal{CN}^{(1)}|$ and $|\textnormal{CN}^{(2)}|$ are significantly large, another words, some nodes within the network can however be influenced by more than one seed. According to this fact and the similarity of seed sets obtained by high-degree and other measures, we proposed a new centrality measure, DegreeDistance, which would choose high-degree seeds in an appropriate distance of each other.  We then improved this measure by inspecting the distance of the non-seed node of highest degree and seed nodes, and if the distance fell below the distance threshold, which was set as 2 and 3, the number of common neighbors (if applicable) of the node and a single seed in each step would determine whether the node could be a seed or not; we put a threshold $\theta$ for this value which was taken the same as the average degree of each dataset in our experiments. On the other hand, since each node has influence over its neighbors, we considered the influence of its neighbors as a factor to keep or remove the in-question node. The experiments showed that the proposed measures are promising as they outperformed other measures on large-scale networks in terms of maximizing the spread of influence with acceptable running time. \\

From the proposed measures, one may improve other centrality measures in a similar way as well as the semi-local centrality measure \cite{chen2012identifying, gao2014ranking}. Another interesting direction is finding a way to pick one seed from a set of nodes all of equal degree. We investigate DegreeDistance for the distance threshold $\DT \in \{ 2, 3\}$, it might be interesting to study the case of $\DT \geq 4$ theoretically and experimentally, and come up with the best distance threshold possible, though it depends on the type of networks. 

\section*{Acknowledgement}
The authors would like to thank the anonymous reviewers of {\href{http://www.journals.elsevier.com/physica-a-statistical-mechanics-and-its-applications/}{Physica A}} for their helpful comments and suggestions. A. Shokrollahi thanks Aaron Clauset for his useful discussion on the Pearson's correlation.


\bibliographystyle{elsarticle-num}
\bibliography{ref1}

\end{document}